\input harvmac.tex
\noblackbox
\input epsf


\input amssym.def
\input amssym.tex

\def\bba{{\bf{A}}}

\def\bbc{{\bf{C}}}

\def\bbe{{\bf{E}}}

\def\bbg{{\bf{G}}}
\def\bbh{{\bf{H}}}

\def\bbk{{\bf{K}}}

\def\bbm{{\bf{M}}}
\def\bbn{{\bf{N}}}
\def\bbp{{\bf{P}}}

\def\bbs{{\bf{S}}}



\input amssym.def
\input amssym.tex
\def\bba{{\Bbb{A}}}

\def\bbc{{\Bbb{C}}}

\def\bbe{{\Bbb{E}}}

\def\bbg{{\Bbb{G}}}
\def\bbh{{\Bbb{H}}}

\def\bbk{{\Bbb{K}}}

\def\bbm{{\Bbb{M}}}
\def\bbn{{\Bbb{N}}}
\def\bbo{{\Bbb{O}}}
\def\bbp{{\Bbb{P}}}

\def\bbs{{\Bbb{S}}}

\def\bbz{{\Bbb{Z}}}




\def\rmk#1{\bigskip\noindent{\bf Remarks} }


\def\figin{\epsfcheck\figin}\def\figins{\epsfcheck\figins}
\def\epsfcheck{\ifx\epsfbox\UnDeFiNeD
\message{(NO epsf.tex, FIGURES WILL BE IGNORED)}
\gdef\figin##1{\vskip2in}\gdef\figins##1{\hskip.5in}
\else\message{(FIGURES WILL BE INCLUDED)}%
\gdef\figin##1{##1}\gdef\figins##1{##1}\fi}
\def\DefWarn#1{}
\def\figinsert{\goodbreak\midinsert}
\def\ifig#1#2#3{\DefWarn#1\xdef#1{fig.~\the\figno}
\writedef{#1\leftbracket fig.\noexpand~\the\figno}%
\figinsert\figin{\centerline{#3}}\medskip\centerline{\vbox{\baselineskip12pt
\advance\hsize by -1truein\noindent\footnotefont{\bf Fig.~\the\figno:} #2}}
\bigskip\endinsert\global\advance\figno by1}


\def\boxit#1{\vbox{\hrule\hbox{\vrule\kern8pt
\vbox{\hbox{\kern8pt}\hbox{\vbox{#1}}\hbox{\kern8pt}}
\kern8pt\vrule}\hrule}}
\def\mathboxit#1{\vbox{\hrule\hbox{\vrule\kern8pt\vbox{\kern8pt
\hbox{$\displaystyle #1$}\kern8pt}\kern8pt\vrule}\hrule}}

%



\def\unlockat{\catcode`\@=11}
\def\lockat{\catcode`\@=12}

\unlockat

\def\newsec#1{\global\advance\secno by1\message{(\the\secno. #1)}
\global\subsecno=0\global\subsubsecno=0\eqnres@t\noindent
{\bf\the\secno. #1}
\writetoca{{\secsym} {#1}}\par\nobreak\medskip\nobreak}
\global\newcount\subsecno \global\subsecno=0
\def\subsec#1{\global\advance\subsecno
by1\message{(\secsym\the\subsecno. #1)}
\ifnum\lastpenalty>9000\else\bigbreak\fi\global\subsubsecno=0
\noindent{\it\secsym\the\subsecno. #1}
\writetoca{\string\quad {\secsym\the\subsecno.} {#1}}
\par\nobreak\medskip\nobreak}
\global\newcount\subsubsecno \global\subsubsecno=0
\def\subsubsec#1{\global\advance\subsubsecno by1
\message{(\secsym\the\subsecno.\the\subsubsecno. #1)}
\ifnum\lastpenalty>9000\else\bigbreak\fi
\noindent\quad{\secsym\the\subsecno.\the\subsubsecno.}{#1}
\writetoca{\string\qquad{\secsym\the\subsecno.\the\subsubsecno.}{#1}}
\par\nobreak\medskip\nobreak}

\def\subsubseclab#1{\DefWarn#1\xdef
#1{\noexpand\hyperref{}{subsubsection}%
{\secsym\the\subsecno.\the\subsubsecno}%
{\secsym\the\subsecno.\the\subsubsecno}}%
\writedef{#1\leftbracket#1}\wrlabeL{#1=#1}}
\lockat

\def\IL{\relax{\rm I\kern-.18em L}}
\def\IH{\relax{\rm I\kern-.18em H}}
\def\IR{\relax{\rm I\kern-.18em R}}
\def\IC{\relax\hbox{$\inbar\kern-.3em{\rm C}$}}
\def\IZ{\relax\ifmmode\mathchoice
{\hbox{\cmss Z\kern-.4em Z}}{\hbox{\cmss Z\kern-.4em Z}}
{\lower.9pt\hbox{\cmsss Z\kern-.4em Z}}
{\lower1.2pt\hbox{\cmsss Z\kern-.4em Z}}\else{\cmss Z\kern-.4em
Z}\fi}
\def\CM {{\cal M}}
\def\CN {{\cal N}}
\def\CR {{\cal R}}
\def\CD {{\cal D}}
\def\CF {{\cal F}}

\def\CL {{\cal L}}

\def\CO {{\cal O}}

\def\CH {{\cal H}}

\def\CS {{\cal S}}

\def\CM {{\cal M}}
\def\CN {{\cal N}}

\def\CO {{\cal O}}

\def\CS {{\cal S }}

\def\diff{{\rm diff}}

\def\Det{{\rm Det}}

\def\wb {\bar{w}}
\font\manual=manfnt \def\dbend{\lower3.5pt\hbox{\manual\char127}}

\def\IZ{\relax\ifmmode\mathchoice
{\hbox{\cmss Z\kern-.4em Z}}{\hbox{\cmss Z\kern-.4em Z}}
{\lower.9pt\hbox{\cmsss Z\kern-.4em Z}}
{\lower1.2pt\hbox{\cmsss Z\kern-.4em Z}}\else{\cmss Z\kern-.4em
Z}\fi}
\def\half {{1\over 2}}
\def\sdtimes{\mathbin{\hbox{\hskip2pt\vrule height 4.1pt depth -.3pt
width
.25pt
\hskip-2pt$\times$}}}
\def\p{\partial}
\def\pb{\bar{\partial}}

\def\CM {{\cal M}}
\def\CN {{\cal N}}

\def\CO {{\cal O}}

\def\CS {{\cal S }}

\def\Det{{\rm Det}}


\def\IZ{\relax\ifmmode\mathchoice
{\hbox{\cmss Z\kern-.4em Z}}{\hbox{\cmss Z\kern-.4em Z}}
{\lower.9pt\hbox{\cmsss Z\kern-.4em Z}}
{\lower1.2pt\hbox{\cmsss Z\kern-.4em Z}}\else{\cmss Z\kern-.4em
Z}\fi}
\def\IA{\relax{\rm I\kern-.18em A}}
\def\IB{\relax{\rm I\kern-.18em B}}
\def\IC{{\relax\hbox{$\inbar\kern-.3em{\rm C}$}}}
\def\ID{\relax{\rm I\kern-.18em D}}
\def\IE{\relax{\rm I\kern-.18em E}}
\def\IF{\relax{\rm I\kern-.18em F}}
\def\IG{\relax\hbox{$\inbar\kern-.3em{\rm G}$}}
\def\IGa{\relax\hbox{${\rm I}\kern-.18em\Gamma$}}
\def\IH{\relax{\rm I\kern-.18em H}}
\def\II{\relax{\rm I\kern-.18em I}}
\def\IJ{\relax{\rm I\kern-.18em J}}
\def\IK{\relax{\rm I\kern-.18em K}}
\def\IL{\relax{\rm I\kern-.18em L}}
\def\IM{\relax{\rm I\kern-.18em M}}
\def\IN{\relax{\rm I\kern-.18em N}}
\def\IO{\relax{\rm I\kern-.18em O}}
\def\IP{\relax{\rm I\kern-.18em P}}
\def\IQ{\relax\hbox{$\inbar\kern-.3em{\rm Q}$}}
\def\IR{\relax{\rm I\kern-.18em R}}
\def\IW{\relax\hbox{$\inbar\kern-.3em{\rm W}$}}

\def\im{{\rm Im}}

\def\inbar{\,\vrule height1.5ex width.4pt depth0pt}

\def\p{\partial}
\def\pb{{\bar \p}}

\font\cmss=cmss10 \font\cmsss=cmss10 at 7pt
\def\IR{\relax{\rm I\kern-.18em R}}

\def\sdtimes{\mathbin{\hbox{\hskip2pt\vrule
height 4.1pt depth -.3pt width .25pt\hskip-2pt$\times$}}}

\def\vol{{\rm vol}}

\def\wb{{\bar{w}}}


\def\boxit#1{\vbox{\hrule\hbox{\vrule\kern8pt
\vbox{\hbox{\kern8pt}\hbox{\vbox{#1}}\hbox{\kern8pt}}
\kern8pt\vrule}\hrule}}
\def\mathboxit#1{\vbox{\hrule\hbox{\vrule\kern8pt\vbox{\kern8pt
\hbox{$\displaystyle #1$}\kern8pt}\kern8pt\vrule}\hrule}}


\def\inbar{\,\vrule height1.5ex width.4pt depth0pt}

\def\p{\partial}

\def\pb{{\bar \p}}

\font\cmss=cmss10 \font\cmsss=cmss10 at 7pt
\def\IR{\relax{\rm I\kern-.18em R}}

\def\sdtimes{\mathbin{\hbox{\hskip2pt\vrule
height 4.1pt depth -.3pt width .25pt\hskip-2pt$\times$}}}

\def\vol{{\rm vol}}

\def\wb{{\bar{w}}}


\def\makeblankbox#1#2{\hbox{\lower\dp0\vbox{\hidehrule{#1}{#2}%
   \kern -#1
   \hbox to \wd0{\hidevrule{#1}{#2}%
      \raise\ht0\vbox to #1{}
      \lower\dp0\vtop to #1{}
      \hfil\hidevrule{#2}{#1}}%
   \kern-#1\hidehrule{#2}{#1}}}%
}%
\def\hidehrule#1#2{\kern-#1\hrule height#1 depth#2 \kern-#2}%
\def\hidevrule#1#2{\kern-#1{\dimen0=#1\advance\dimen0 by #2\vrule
    width\dimen0}\kern-#2}%
\def\openbox{\ht0=1.2mm \dp0=1.2mm \wd0=2.4mm  \raise 2.75pt
\makeblankbox {.25pt} {.25pt}  }
\def\opensquare{\ht0=3.4mm \dp0=3.4mm \wd0=6.8mm  \raise 2.7pt \makeblankbox
{.25pt} {.25pt}  }

\def\sector#1#2{\ {\scriptstyle #1}\hskip 1mm
\mathop{\opensquare}\limits_{\lower 1mm\hbox{$\scriptstyle#2$}}\hskip 1mm}

\def\tsector#1#2{\ {\scriptstyle #1}\hskip 1mm
\mathop{\opensquare}\limits_{\lower 1mm\hbox{$\scriptstyle#2$}}^\sim\hskip 1mm}

\lref\acharyaii{B. S. Acharya,
``N=1 Heterotic/M-theory Duality and Joyce Manifolds,''
hep-th/9603033; Nucl.Phys. B475 (1996) 579-596}

\lref\lagew{L. Alvarez-Gaum\'e and E. Witten,
``Gravitational Anomalies,'' Nuc. Phys. {\bf B234}(1983),
269}
\lref\aft{I. Antoniadis, S. Ferrara, and T. Taylor, ``${\CN}=2$ heterotic
superstring
and its dual theory in five dimensions," hep-th/9511108, Nucl. Phys. {\bf B
460} (1996) 489.}

\lref\apostol{T. Apostol, {\it Modular Functions and
Dirichlet Series in Number Theory}, Springer Verlag 1990}
\lref\spinning{Breckenridge et. al. Spinning
black holes}
\lref\dmvv{DMVV}
\lref\ez{Eicher and Zagier}
\lref\tkawai{T. Kawai}
\lref\kawai{T. Kawai, Y. Yamada, and S. -K. Yang,
``Elliptic genera and $N=2$ superconformal
field theory,'' hepth/9306096 }
\lref\maldacenai{J. Maldacena, ``Black holes and
D-branes,'' hep-th/9705078}
\lref\sv{Strominger and Vafa}
\lref\acharya{B. Acharya,
`` M Theory, Joyce Orbifolds and Super Yang-Mills,''
hep-th/9812205.}
\lref\ads{I. Affleck, M. Dine, and N. Seiberg,
``Dynamical supersymmetry breaking in
supersymmetric QCD,'' Nucl. Phys.
{\bf 241B}(1984) 493.}
\lref\ahw{I. Affleck, J. Harvey, and E. Witten,
Nucl. Phys. {\bf B206} (1982) 413.}
\lref\agdpm{L. Alvarez-Gaum\'e, S. Della Pietra, and G. Moore,
``Anomalies and odd dimensions,'' Ann. Phys.
{\bf 163}(1985)288.}
\lref\aspinwall{P. Aspinwall, ``K3 surfaces and
string duality,'' hep-th/9611137}
\lref\agm{P. Aspinwall, B. Greene, and D. Morrison,
``The monomial divisor mirror map,''
 Internat. Math. Res. Notices (1993), 319-337; alg-geom/9309007.}
\lref\baggerwitten{J. Bagger and E. Witten,
``Quantization of Newton's Constant in
Certain Supergravity Theories,'' Phys. Lett.
{\bf 115B}(1982) 202}
\lref\embedi{I. Bandos, P. Pasti,  D. Sorokin,  M. Tonin, D. Volkov,
``Superstrings and supermembranes in the
doubly supersymmetric geometrical approach,''
Nucl. Phys.{\bf B446} (1995) 79; hep-th/9501113. }
\lref\bd{T. Banks and M. Dine, ``Couplings and
Scales in Strongly Coupled Heterotic
String Theory,'' hep-th/9605136; hep-th/9609046.}
\lref\bbs{K. Becker, M. Becker and A. Strominger,
``Fivebranes, membranes and non perturbative
string theory,''
Nucl. Phys. {\bf 456B} 130;  hep-th/9507158. }
\lref\bbmo{K. Becker, M. Becker, D. Morrison, H. Ooguri, Y. Oz,
and  Z. Yin, ``Supersymmetric cycles in exceptional
holonomy manifolds and Calabi-Yau 4 folds,''
 Nucl.Phys.{\bf B480} (1996) 225; hep-th/9608116.}
\lref\bbii{K. Becker and M. Becker,
`` Instanton Action for Type II Hypermultiplets,''
hep-th/9901126. }

\lref\bst{E. Bergshoeff, E. Sezgin, and P.K. Townsend,
``Supermembranes and 11-dimensional supergravity,''
Phys. Lett. {\bf 189B}(1987)75;
``Properties of the Eleven-Dimensional Supermembrane
Theory,'' Ann. Phys. {\bf 185}(1988)330.}
\lref\bsv{M. Bershadsky, V. Sadov, and
C. Vafa, ``D-Branes and Topological Field Theories,''
Nucl.Phys. B463 (1996) 420-434; hep-th/9511222.}
\lref\bbrt{D. Birmingham, M. Blau, M.
Rakowski, and G. Thompson,
``Topological Field Theories,'' Phys. Reports {\bf 209}(1991)129. }
\lref\bryant{R.L. Bryant, ``Metrics with exceptional
holonomy,'' Ann. Math. {\bf 126}(1987) 525.}
\lref\cmrtft{
S. Cordes, G. Moore and S. Rangoolam,  ``Lectures on  2D
Yang-Mills Theory, Equivariant Cohomology and Topological Field Theory",
hep-th/9411210. }
\lref\elevss{E. Cremmer and S. Ferrara,
``Formulation of 11-dimensional supergravity in
superspace,'' Phys. Lett. {\bf 91B}(1980) 61;
L. Brink and P. Howe, ``Eleven-dimensional supergravity
on the mass shell in superspace,'' Phys. Lett.
{\bf 91B}(1980)384.}
\lref\cfgv{E. Cremmer, S. Ferrara, L. Girardello,
and A. Van Proeyen, ``Yang-Mills theories with
local supersymmetry: Lagrangian, transformation
laws and super-Higgs effect,'' Nucl. Phys. {\bf B212}
(1983) 413.}
\lref\daifreed{X. Dai and D.S. Freed,
``$\eta$-invariants and determinant lines,''
J. Math. Phys. {\bf 35} (1994) 5155-5194}
\lref\dpp{B. de Wit, K. Peeters, and J. Plefka,
``Superspace geometry for supermembrane
backgrounds,'' Nucl. Phys. {\bf B532} (1998) 99; hep-th/9803209.}
\lref\dineseib{M. Dine and N. Seiberg,
``Is the superstring semiclassical?'' in
{\it Unified String Theories}, M. Green and D. Gross eds.
World Scientific 1986.}
\lref\dsww{M. Dine, N. Seiberg, X.-G. Wen and
E. Witten, ``Nonperturbative effects on the string
world sheet I,II'' Nucl. Phys. {\bf B278}(1986)769;
Nucl. Phys. {\bf B289}(1987)319.}
\lref\distlerkachru{J. Distler and S. Kachru,
``(0,2) Landau-Ginzburg Theory,''
hep-th/9309110; Nucl. Phys. {\bf B413}
(1994)213.}
\lref\dontom{S.K. Donaldson and R.P. Thomas,
``Gauge theory in higher dimensions,'' in  The geometric universe (Oxford,
1996), 31--47, Oxford Univ. Press, Oxford, 1998}
\lref\fhmm{D. Freed, J. Harvey, R. Minasian,
and G. Moore,
``Gravitational Anomaly Cancellation for M-Theory Fivebranes,''
Adv.Theor.Math.Phys.{\bf 2} (1998) 601-618;hep-th/9803205.}
\lref\gpp{G. Gibbons, D. Page, and C. Pope,
``Einstein metrics on $S^3$, $R^3$, and
$R^4$ bundles,'' Commun. Math. Phys.
{\bf 127} (1990) 529.}
\lref\ghm{R. Gregory, J.A. Harvey, and G. Moore,
``Unwinding strings and T duality of Kaluza-Klein
and H monopoles,'' hep-th/9708086;
 Adv.Theor.Math.Phys.1:283-297,1997}
\lref\hls{ J.A. Harvey, D.A. Lowe, A. Strominger,
`` N=1 String Duality,''  hep-th/9507168,
 Phys.Lett. B362 (1995) 65-72.}

\lref\hl{R. Harvey and H.B. Lawson Jr.,
``Calibrated Geometries,'' Act. Math. {\bf 148} (1982) 47.}
\lref\hitchin{N.J. Hitchin,
``The moduli space of special Lagrangian submanifolds,''
dg-ga/9711002.}
\lref\horavawitten{
P. Horava and E. Witten,
``Heterotic and Type I String Dynamics from Eleven Dimensions,''
hep-th/9510209, Nucl.Phys. B460 (1996) 506-524;
``Eleven-Dimensional Supergravity on a Manifold with Boundary,''
hep-th/9603142, Nucl.Phys. B475 (1996) 94-114}
 \lref\horava{P. Horava, ``Gluino condensation in
strongly coupled heterotic string theory,'' Phys. Rev. {\bf D54}
(1996) 7561;
hep-th/9608019}
\lref\hkty{ S. Hosono, A. Klemm, S. Theisen and S.-T. Yau,
``Mirror Symmetry, Mirror Map and Applications to Complete Intersection
Calabi-Yau Spaces, ''
hep-th/9406055;Nucl.Phys. B433 (1995) 501-554}
\lref\embedii{Howe et. al. superembeddings}
\lref\hubsch{T. Hubsch, {\it Calabi Yau Manifolds:
A Bestiary for Physicists}, World Scientific, 1992.}
\lref\joycei{D. Joyce, ``Compact Riemannian 7-manifolds
with holonomy $G_2$ Parts I, II'' J. Diff. Geom.
{\bf 43} (1996) 291.}
\lref\joyceii{D. Joyce,  Inv. Math. ICM proceedings }
\lref\kontsevich{M. Kontsevich,
``Homological Algebra of Mirror
Symmetry,'' Proc. of the 1994 International
 Congress of Mathematicians,
p.120, Birkh\"auser, 1995; alg-geom/9411018.}
\lref\losev{L. Baulieu,
A. Losev, and N. Nekrasov,
``Chern-Simons and twisted supersymmetry in
various dimensions,''
hep-th/9707174; Nucl.Phys.B522:82-104,1998}
\lref\michlaw{H.B. Lawson and M-L Michelsohn,
{\it Spin Geometry}, Princeton University
Press, 1989}
\lref\msw{J. Maldacena, A. Strominger and E. Witten, ``Black Hole Entropy
in M Theory,'' J. High Energy Phys. 12 (1997) 2;, hep-th/9711053.}
\lref\mmt{R. Minasian, G. Moore, and
D. Tsimpis, ``Calabi-Yau black holes and (0,4) sigma models,''
hep-th/9904217.}
\lref\mclean{R.C. McLean, ``Deformations of
calibrated submanifolds,''  Commun. Anal. Geom.
{\bf 6}(1998) 705. }
\lref\neklaw{A. Lawrence and N. Nekrasov,
``Instanton sums and five-dimensional gauge theories,'' hep-th/9706025;
Nucl.Phys. B513 (1998) 239-265}
\lref\morrpless{D.R. Morrison and M.R. Plesser,
``Towards mirror symmetry as duality for
two-dimensional abelian gauge theories,''
hep-th/9508107; Nucl.Phys.Proc.Suppl.46:177-186,1996}
 \lref\ovrutii{A. Lukas, B.A. Ovrut, and D. Waldram,
``Five-Branes and Supersymmetry Breaking in
M-Theory,'' hep-th/9901017.}
\lref\paptown{G. Papadopoulos and Townsend,
``Compactifications of D=11 supergravity on
spaces of exceptional holonomy,'' Phys. Lett. {\bf B357}
(1995) 300;  hep-th/9506150.}
\lref\polishchuk{A. Polishchuk and  E. Zaslow,
``Categorical Mirror Symmetry: The Elliptic Curve,''
math.AG/9801119; D. Arinkin, A. Polishchuk.
``Fukaya category and Fourier transform,''  math.AG/9811023.}
\lref\redlich{ A.N. Redlich, ``Gauge noninvariance
and parity violation of three dimensional fermions,''
Phys.Rev.Lett.{\bf 52} (1984) 18.}
\lref\rw{L. Rozansky and E. Witten,
``Hyperk\"ahler geometry and three manifold
invariants," hep-th/9612126.}
\lref\swrec{N. Seiberg and E. Witten,
``The D1/D5 system and singular CFT,''
hep-th/9903224; JHEP 9904:017,1999}

\lref\shati{S. Shatashvili and C. Vafa,
``Superstrings and manifolds of exceptional
holonomy,'' hep-th/9407025; Selecta Mathematica,
{\bf 1}(1995)347; ``Exceptional Magic,'' Nucl. Phys.
Proc. Supp.  {\bf 41}(1995)345}
\lref\shatii{M. Green and S. Shatashvili,
``Notes on Compactification of string
theory to two and three dimensions,''
unpublished.}
\lref\shatiii{B. Freivogel, M. Lippert, and
S. Shatashvili, to appear}
\lref\shibusa{Y. Shibusa,
``11-dimensional curved backgrounds for supermembrane in superspace,''
hep-th/9905071}
\lref\silverwitten{E. Silverstein and E. Witten,
`` Criteria for conformal invariance of (0,2) models,''
hep-th/9503212; Nucl.Phys.B444:161-190,1995}
\lref\singer{I.M. Singer, ``Families of Dirac operators
with applications to physics,'' in
{\it The Mathematical Heritage of Elie Cartan},
Asterisque (1985) 323}
\lref\stromwitten{A. Strominger and E. Witten,
``New manifolds for superstring compactification,''
Commun.Math.Phys.{\bf 101} (1985) 341.}
\lref\syz{A. Strominger, S.-T. Yau, and
E. Zaslow, ``Mirror symmetry is T-duality,''
Nucl.Phys.{\bf B479} (1996) 243; hep-th/9606040.}
\lref\thomas{R.P. Thomas,
``A holomorphic Casson invariant
 for Calabi-Yau 3-folds, and bundles on K3 fibrations,''
math.AG/9806111.}
\lref\townrev{P.K. Townsend,
``The eleven-dimensional supermembrane
revisited,'' hep-th/9501068; Phys.Lett.B350:184-187,1995}

\lref\tyurin{A. Tyurin, ``Special Langrangian geometry and slightly deformed
algebraic geometry (spLag and sdAG),'' math.AG/9806006.}
\lref\townsend{P. Townsend,
``D-branes from M-branes,''
 Phys.Lett.{\bf B373} (1996) 68; hep-th/9512062.}
\lref\wb{J. Wess and J. Bagger, {\it Supersymmetry
and Supergravity}, Second ed., Princeton University
Press 1992.}
\lref\jones{E. Witten, ``Quantum Field theory and the
jones polynomial,'' Commun.Math.Phys.{\bf 121} (1989) 351.}
\lref\tqft{E. Witten,
``Topological Quantum Field Theory,''
Commun. Math. Phys. {\bf 117} (1988)
353.}
\lref\cohoft{E. Witten, Introduction to Cohomological
Field Theory,  in Trieste Quantum Field Theory 1990:15-32}

\lref\ewglsm{E. Witten, ``Phases of N=2 Theories
in Two Dimensions,'' hep-th/9301042;
Nucl. Phys. {\bf B403}(1993)159.}
\lref\fourflux{E. Witten, ``On flux quantization in
$M$-theory and the effective action,''
hep-th/9609122; J.Geom.Phys. 22 (1997) 1-13 }
\lref\wittga{E. Witten, ``Global gravitational
anomalies,'' Commun.Math.Phys.{\bf 100} (1985) 197.}
\lref\abmodels{E. Witten, ``Mirror manifolds
and topological field theory,'' hep-th/9112056
}
\lref\wfv{E. Witten, ``Non-perturbative superpotentials
in string theory,'' hep-th/9604030;
Nucl.Phys.{\bf B474} (1996) 343.}
\lref\wittfive{E. Witten, ``Five-Brane Effective Action In M-Theory,''
J. Geom. Phys. {\bf 22} (1997) 103;
hep-th/9610234.}
\lref\wittsc{E. Witten,
``Strong coupling expansion of Calabi-Yau
compactifications,'' Nucl.Phys.{\bf B471} (1996) 135; hep-th/9602070.}
\lref\fw{Freed and Witten, ``Anomalies in string
theory with D-branes,''  to appear}
\lref\wittendb{E. Witten, ``Worldsheet
corrections via D-instantons,'' to appear }

\lref\ovrut{R. Donagi, A. Lukas, B.A. Ovrut,
and D. Waldram, ``Nonperturbative vacua and
particle physics in M-theory,'' hep-th/9811168;
``Holomorphic vector bundles and
nonperturbative vacua in M-theory,'' hep-th/9901009;
``Moduli spaces of five-branes on
elliptic Calabi-Yau threefolds,''  hep-th/9904054.}
\lref\eisenhardt{L. Eisenhart, {\it Riemannian Geometry}, ch. IV}
\lref\spivak{M. Spivak, {\it A comprehensive introduction
to differential geometry}, vol. IV, ch. 7.}
\lref\kobayashi{S. Kobayashi and K. Nomizu,
{\it Foundations of Differential Geometry}, vol. II, ch. 7.}
\lref\chern{S.S. Chern, M. Do Carmo and S. Kobayashi,
``Minimal submanifolds of a sphere with second
fundamental form of constant length,'' in {\it Selected Works}.
}
\lref\gvafa{R. Gopakumar and C. Vafa, ``M-Theory and Topological Strings-
I,II,'' hep-th/9809187,hep-th/9812127.}
\lref\bks{L. Baulieu, H. Kanno and I. Singer, ``Special Quantum Field
Theories in Eight and Other Dimensions,''
hep-th/9704167; Commun. Math. Phys. {\bf 194}
(1998) 149.}
\lref\fks{J.M. Figueroa-O'Farrill, C. Kohl and B. Spence,
``Supersymmetric Yang-Mills, Octonionic Instantons, and Triholomorphic
Curves,'' hep-th/9710082;
Nucl. Phys. {\bf B521} (1998) 419.}
\lref\octinst{E. Corrigan, C. Devchand, D. B. Fairlie and J. Nuyts,
Nucl. Phys. {\bf B214} (1983) 452; D. B. Fairlie and J. Nuyts, J. Phys.
A {\bf 17} (1984) 2867; S. Fubini and H. Nicolai, Phys. Lett. {\bf 155B}
(1985) 369.}

\Title{\vbox{\baselineskip12pt
\hbox{EFI-99-22}
\hbox{YCTP-P15-99}
\hbox{IASSNS-99/57}
\hbox{hep-th/9907026}
}}
{\vbox{\centerline{
Superpotentials and Membrane Instantons }
}}

\centerline{Jeffrey A. Harvey}
\medskip
\centerline{Enrico Fermi
Institute and Department of Physics}
\centerline{University of
Chicago, 5640 Ellis Avenue, Chicago, IL 60637}
\bigskip
\centerline{  Gregory Moore}
\medskip
\centerline{Department of Physics, Yale University,}
\centerline{Box 208120, New Haven, CT 06520}

\bigskip
\centerline{\bf Abstract}

\bigskip
We investigate nonperturbative effects in
$M$-theory compactifications arising from
wrapped membranes. In particular, we show
that in $d=4, \CN=1$ compactifications
along manifolds of $G_2$ holonomy, membranes
wrapped on rigid supersymmetric 3-cycles
induce nonzero corrections to the superpotential.
Thus, membrane instantons destabilize many
$M$-theory compactifications. Our computation
shows that the low energy description of
membrane physics is usefully described in
terms of three-dimensional topological
field theories, and the superpotential is
expressed in terms of topological invariants
of the 3-cycle.
We discuss briefly some applications of
these results. For example, using
 mirror symmetry we derive a
counting formula for supersymmetric
three-cycles in certain Calabi-Yau manifolds.

\Date{July 4, 1999}

\newsec{Introduction}

The importance of instanton computations
in string theory and in M-theory can hardly
be overstated. To cite but one reason,
the understanding of instanton
effects is a necessary
ingredient in attempts to make realistic M/string theory
models which address the problems of vacuum selection
and supersymmetry breaking.

Compared to instantons in gauge theory
and string theory, $M$-brane instantons
have not been so thoroughly discussed.
These effects were first
discussed in the fundamental paper of
Becker, Becker, and Strominger \bbs.
There has been some work on non-perturbative
corrections in   compactifications
of $M$-theory with 4 unbroken supersymmetries
beginning with  Witten's investigation
of   five-brane instantons \wfv .
Since $M2$ branes can sometimes be interpreted
in terms of world-sheet instantons
in IIA string theory  (see, e.g. \bbs\aft\neklaw)
a fair amount is known in this case. Nevertheless,
we believe that much remains to be understood
concerning the  general computation of $M$-brane
instantons, and this paper is a modest step in
that direction.

Specifically, in the present paper
 we discuss instanton effects
associated with wrapped $M2$
branes in $M$ theory compactifications to
three and four dimensions. Throughout
most of the paper we focus on the
example of $M$-theory compactification
on smooth seven-manifolds of $G_2$
holonomy (henceforth abbreviated as $G_2$-manifolds).
We will show that in  such compactifications
membrane instantons can induce
 a nonzero superpotential.
As an example, rigid membranes lead
to a contribution to the superpotential
given in  equations (2.13) and (6.10) below.

Here is a brief overview of the paper.
In section 2 we review a few aspects of
Kaluza-Klein reduction of 11-dimensional
supergravity on $G_2$ manifolds.
In section 3 we complain about the
absence of a clear set of rules for
computing $M$-brane instanton effects
(due to the want of a fundamental formulation
of $M$-theory), and outline the practical
procedure we will adopt.

In sections 4 and 5 we consider the
low energy action of a membrane in
curved superspace. The
upshot of our discussion
is that the low energy fluctuations
of a membrane wrapped on a supersymmetric
3-cycle are  usefully thought of in terms of
certain three-dimensional
topological field  theories. It is well-known
that in string theory {\it two-dimensional}
topological sigma models
(the $A-$ and $B-$ models, and their
couplings to topological gravity)
are quite relevant to the computation of
superpotentials
\dsww\abmodels. Not surprisingly,  we are finding
that in $M$-theory an analogous role
is played by three-dimensional topological
field theory.

Section 6 contains the key results of the
paper. We derive
(following the provisional
rules of section 3) the contribution to the
superpotential from a rigid membrane
instanton, and derive one-loop determinants
and zeromodes needed in the more general
case of arbitrary membrane instantons.
We discuss briefly the relation of the
phase of the instanton contribution
to the one-loop determinants, following
a similar discussion by Witten in the
example of D1-brane instantons
\wittendb.

In the remainder of the paper we turn to
examples and applications of our results.
In section 7 we discuss a simple class of
$G_2$ manifolds and show that there do
exist examples of the kinds of effects we have
discussed. (We also give a relevant
example in appendix C.)
Section 8 contains some brief
comments on the extension of this work to compactifications
of IIA string theory on $G_2$ manifolds.
 Section 9 concerns
generating functions counting supersymmetric
3-cycles in Calabi-Yau  3-folds admitting a
real structure.  Section 10
examines consequences for nonperturbative
effects in the heterotic string. Section 11
addresses some effects of open membrane
instantons. Many conventions and notations
can be found in appendices A and B.

Some closely related issues to those
discussed in this paper have been addressed
by B.S.  Acharya in \acharya.
Acharya focuses
on certain interesting singular $G_2$
manifolds. In the present
paper we usually work with the
case of {\it smooth} $G_2$ manifolds.

\newsec{Compactification on $G_2$ manifolds}

The low energy limit of $M$-theory is described
by 11-dimensional supergravity.
This theory consists of
a metric $g $,
a 3-form   $C  $,
and a gravitino $\Psi  $, defined on
 an eleven-dimensional Lorentzian
spin manifold $M_{11}$.
The bosonic action enters the path integral
through
\foot{An explanation of our
conventions for scales and dimensions is in
appendix B. }
\eqn\goodnorm{
\exp\Biggl[  {i \over  (2 \pi)^2 \ell^9} \int  \vol(g)   \CR(g) +
{i \over  (2 \pi)^2 \ell^3} \int   \half d C \wedge * dC
 -{   i \over 6\cdot (2 \pi)^2 } \int_{M_{11} } G \wedge G
\wedge  C     +\cdots \Biggr]
}
where $\ell$ is the eleven-dimensional Planck length  and $G=dC$. The ellipsis
indicates the presence of
higher order terms in
the low energy expansion. These
will be ignored in what follows.
The field strength $G$ is normalized by requiring that
\eqn\normgf{
\int_{S^4} {G\over 2 \pi}  =1
}
where the integral is taken around a 4-sphere
linking the basic five-brane
in   $\IR^{1,10 }$.

In this paper $M_{11}$ will be taken to be smooth, and
all relevant lengthscales are large compared to
$\ell$.
Moreover, we focus on vacua
with smooth Ricci flat direct product metrics:
\eqn\vac{
\eqalign{
M_{11} & = M_4 \times X \cr
g(x,y) &  = g_{\mu\nu}(x)dx^\mu dx^\nu + g_{mn}(y) dy^m dy^n \cr}
}
($0\leq \mu,\nu\leq 3$, $1\leq m,n\leq 7$.)
We take  $X$ to be a manifold of $G_2$ holonomy.
In particular we assume $X$
has a   covariantly constant
spinor $\nabla_m \vartheta = 0$, unique
up to scale. Thus,
compactification on $X$ leads to a
theory with $d=4, \CN=1$ supersymmetry.
When we work in Euclidean signature we will take
$M_4$ to be a hyperk\"ahler manifold.

We complete the specification of the
background by taking
the background value of
$C$ to be a real harmonic 3-form on $X$,
and thus $G=dC =0$.
\foot{We are ignoring an important subtlety here.
The following remarks were explained to us by
E. Witten.
In compactifications of $M$ theory on $X$
the quantization condition is shifted \fourflux\ and becomes
the statement that
\eqn\fluxshift{ \left[ {G \over 2 \pi} \right] - {\lambda \over 2 }
\in H^4(X, \IZ) }
where $[G/2 \pi]$ is the cohomology class of $G/2 \pi$ and
$\lambda (X) = p_1(X)/2$. For a $G_2$-manifold
$\lambda$ is always even, so that it is  consistent to set
$G=0$. The argument
is the following: Note that
$p_1(X) = p_1(X \times S^1)$. Now
$X \times S^1$ is an eight-dimensional spin manifold
and
$\lambda$ is even or odd
on an eight-dimensional spin manifold
according to whether the intersection form is
even or odd \fourflux. Now
 observe that the intersection form of $X \times S^1$
  is obviously even.  }

The Kaluza-Klein reduction of $M$-theory
on $X$ is completely straightforward.
The field multiplets were worked out in
\paptown. The massless
multiplets consist of
the supergravity multiplet $(g_{\mu\nu}, \psi_\mu)$,
chiral multiplets $(z^i, \chi^i)$, $i = 1, \dots, b_3(X)$,
and  $U(1)$ vector multiplets $(\lambda^I, A_\mu^I)$,
$I=1, \dots, b_2(X)$.

The above multiplets may be derived as follows.
The $U(1)$ gauge fields are obtained from
the 3-form $C$. Choosing a basis of harmonic
forms
$\{ \omega^{(2)}_I\}_{I=1,\dots, b_2(X)}$
for $\CH^2(X;\IZ)$ and a  basis
$  \{ \omega^{(3)}_i\}_{i=1,\dots, b_3(X)}   $
for $\CH^3(X;\IZ) $
we write the Kaluza-Klein ansatz
\eqn\thrfrm{
C=\sum_{I=1}^{b_2(X)} A^I(x) \wedge \omega^{(2)}_I(y)
+ \ell^{-3} \sum_{i=1}^{b_3(X)} P^i(x)  \omega^{(3)}_i(y)
}
where the $P^i$ are pseudo-scalars.
Using the normalization \normgf\ large
gauge transformations form a maximal rank
discrete
subgroup of $2 \pi \CH^2(X;\IZ)$. Thus the
gauge group is a finite cover of the
torus  $H^2(X;\IR)/(2\pi H^2(X;\IZ))$.

Let us now consider the chiral multiplets.
Using the covariantly constant spinor
one can construct  a ``$G_2$ calibration,'' i.e., a
covariantly constant 3-form:
$\Phi_{mnp} = \bar \vartheta \gamma_{mnp} \vartheta$.
In an appropriate local orthonormal frame the components of
$\Phi$ are the structure constants of the
octonions $\bbo$.
Conversely, such a covariantly constant
3-form can be used to characterize the $G_2$
structure \refs{\gpp \joycei}.
\foot{See  particularly \joycei, Theorem C.
Much useful background material can be
found in \bryant\shati.}
We may thus define real scalar fields
$S^i(x)$ by
choosing a base point $G_2$ structure $\Phi_0$
and associating the $G_2$ metric to a torsion
free calibration
\eqn\defessi{
\Phi = \Phi_0 + \ell^{-3}  \sum_{i=1}^{b_3(X)}
S^i(x) \omega_i^{(3)}(y)
}

By straightforward reduction of the supergravity
action (described below)
one finds that the K\"ahler target space
has holomorphic tangent space
$ T^{1,0} \CM   = H^3(X ;\IR) \oplus H^3(X;\IR) $,
with the obvious complex structure.
The holomorphic coordinates on this
space are given by
\eqn\chiralsc{
C  + i\ell^{-3} \Phi =  i \Phi_0 +\ell^{-3}  \sum_{i=1}^{b_3(X)}
z^i(x) \omega_i^{(3)}(y)
}
with $z^i = P^i +i S^i$.
This is the analog for $G_2$ manifolds of the
complexified K\"ahler class familiar from
string  compactification
on Calabi-Yau 3-folds.
 Holomorphy in
 $C+ i \ell^{-3} \Phi$ will play an important
role in the discussion below.

The
Kaluza-Klein expansion of the gravitino is
\eqn\fermkk{
\eqalign{
dx^M \Psi_M & = dx^\mu \Psi_\mu + dx^m \Psi_m \cr
\Psi_{\mu}(x,y) & = \psi_\mu(x) \otimes \vartheta(y) \cr
\Psi_m(x,y) & = \ell^3 \sum_{i=1}^{b_3}   \omega^{(3)}_{i,mpq}(y) \Gamma^{pq}
\chi^i(x) \otimes \vartheta(y) \cr
& + \ell^2 \sum_{I=1}^{b_2}   \omega^{(2)}_{I,mp}(y) \Gamma^{p}
\lambda^I(x)  \otimes \vartheta(y) + \cdots \cr}
}

Now let us consider the low energy effective
action. The   effective action at the two-derivative
order is determined
by a choice of a K\"ahler target space $\CM$
for the chiral scalars   $z^i$,  a holomorphic
gauge kinetic function $\tau_{IJ}(z)$,
and a holomorphic superpotential $W(z)$.
\foot{The term $\int C \wedge I_8$ {\it does}
contribute, but at 4th order in derivatives.}

The gauge kinetic term, entering
the 4D action through (among other terms)
\eqn\normtau{
{1 \over  32 \pi} \int_{M_4} \im \biggl[
\tau_{IJ} (F^I - i \tilde F^I) (F^J - i \tilde F^J)
\biggr],
}
comes from the kinetic terms and Chern-Simons
term for $C$. It is given by
\eqn\gaugekn{
\tau_{IJ}    = {2 \over  \pi} \int_X (C +i  \ell^{-3} \Phi) \wedge
\omega^{(2)}_I\wedge\omega^{(2)}_J.
}
Harmonic forms on $X$ transform in the
${\bf 14}$ of the holonomy group and
therefore satisfy
$* \omega^{(2)}_I=- \Phi \wedge \omega^{(2)}_I $.
It follows that  $\im \tau_{IJ}$ is definite
when $X$ is smooth\foot{Again, the subtlety mentioned in
footnote 2 above is important at this point.
The axion terms associated with $C$ are
only really well-defined when combined
with the one-loop Rarita-Schwinger determinant,
as in \fourflux. }.

The  K\"ahler potential can be deduced by
looking at the kinetic terms of the scalars $P^i$.
Using identities similar to the Calabi-Yau case one
obtains:
\eqn\kahlpot{
K = - \log \biggl[ {\int_X \sqrt{\det g} \over  \int_X \sqrt{\det g_0}
}\biggr]
}
where $g_0$ corresponds to the basepoint $\Phi_0$. We have chosen the additive
constant in $K$ so that the
4D Newton constant is $ \pi \ell^9/(4\int_X \sqrt{\det g_0}) $.

%
%

The superpotential is zero in the low energy
supergravity approximation. This is proved
using an argument analogous to that in
\dsww. As shown in  \refs{\baggerwitten,\wb}
 the superpotential  $W$ is a global
holomorphic section of a  negative Hermitian line bundle
$\CL \rightarrow \CM$ with
\eqn\linebdl{
c_1(\CL) =  - {i \over  2 \pi   } \p \pb \log \parallel W \parallel^2
:=- {i \over  2 \pi   } \p \pb \log e^K \vert
 W \vert^2  .
}
In the absence of membrane
instantons there is a continuous Peccei-Quinn  symmetry
$C \rightarrow C + \omega$, $\omega\in \CH^3(M_{11};\IR)$.
On the other hand, the overall volume
$\vol(X)$ is equal to ${1 \over  7} \int \Phi \wedge *
\Phi$, so the energy expansion is an
expansion in inverse powers of $\im (C + i \ell^{-3} \Phi)$.
Now we use holomorphy of $W$  to conclude that
there are no corrections.

On the other hand,
membrane instantons
from   Euclidean M2 branes
wrapping   3-cycles $\Sigma\subset X$ violate
the PQ symmetry, breaking it to some
discrete subgroup of $\CH^3(M_{11};\IZ)$.
Thus membrane instantons   can
induce a superpotential. We will
show that indeed the contribution to the
superpotential of  a {\it rigid}
supersymmetric rational homology
sphere in the $G_2$ manifold
is
\eqn\superp{
\Delta W \propto
\vert H_1(X;\IZ) \vert \exp[ i \int_{\Sigma}(  C + i \Phi) ]
}

The proportionality constant
(independent of $C, \Phi$ ) is   positive
and  associated with bulk supergravity
determinants. A more precise result for this
constant awaits the solution of some conceptual
problems described in the next section.

We should also worry about 5-brane instanton
effects. In \joycei\ (II. Proposition 1.1.1)
Joyce shows that $\pi_1(X)$ (and hence
$H_6(X;\IZ)$ ) is finite if
the holonomy is all of $G_2$. If the holonomy
group is smaller the group $H_6(X;\IZ)$ might or might not
be  finite. Even when it is
finite there might be interesting effects.
However, generically, there will be no
5-brane instantons.

\newsec{On the rules for computing  membrane
instanton effects }

\subsec{Divertimento}\foot{With apologies
to S. Coleman, and G. Galilei. }

\noindent
{\bf Salviati:} Sagredo! When are we going
to write that paper on membrane instantons!? Basta! Basta!

\noindent
{\bf Sagredo:} Salviati! How can we write a
paper on M-theory instantons when we don't understand the fundamental
formulation of M-theory?!

\noindent
{\bf Salviati:} Piano, piano. Look.
We have now learned the lesson
of $p$-brane democracy. $M$-theory
is a theory of fundamental M2-branes,
or, by duality, a theory of
fundamental M5-branes.
Eleven-dimensional supergravity
is just a collective excitation, as in string theory.
Indeed, if we study the membrane solution in supergravity
we find a {\it timelike} singularity in the metric!
Thus, as for the string solution of ten-dimensional
supergravity,  the membrane is fundamental. Therefore,
we  simply
need to follow the obvious generalization of the rules of
string theory and reduce the computation of fermion bilinears
in spacetime to a computation of the correlation function of
vertex operators in the M2-brane theory. The gravitino vertex
operator has even been computed for us in \bbs.

\noindent
{\bf Sagredo:} No. I cannot agree. There is no
``obvious generalization of the rules of string
theory.''
 No one has found the graviton by quantizing
membranes. On the other hand, Matrix Theory {\it is} a proposal
for a fundamental formulation. Furthermore,
 in matrix theory the M2-brane {\it is}  a collective
excitation composed of D0-branes.  So
let's do the computation directly in Matrix Theory!

\noindent
{\bf Salviati:} For membranes wrapped on
supersymmetric cycles in $G_2$ manifolds...?
Prego: ``to appear'' or ``in preparation'' ?

\noindent
{\bf Sagredo:} O.K. Perhaps this is not necessary.
Still,   membranes are not fundamental,
but simply collective  excitations of  some fundamental
degrees of freedom in some fundamental
formulation of M theory. The best we can do is take
the supergravity soliton description seriously and
think of the membrane modes  as collective coordinates
in the solitonic description. The computation of
membrane instantons indeed involves a path integral
of a 3D field theory, but this should simply
be regarded as an integral over
 collective coordinates. After all,
the saddle point
technique is just a way of reducing the number of
integrals you have to do, and we are doing nothing
more than a saddle point approximation to the
``path integral for $M$-theory'' (if that is even the right formulation).
The determinants of the saddle-point approximation
are just the 11D supergravity determinants in the
membrane solution background.

\noindent
{\bf Salviati:} Alas, Sagredo, I fear you have erred in
two ways. First, recall that in
't Hooft's calculation of instanton effects in Yang-Mills
theory the space of collective coordinates is  finite
dimensional, and the action is constant on this space of
collective coordinates.
Do you really want to identify all of the
superembeddings  $(X(s), \Theta(s))$ of the membrane
as collective coordinates? This cannot be, for the
action is not constant on this space.
Second, the singularity in the membrane metric makes
the supergravity determinants you propose to calculate
ill-defined. One needs to specify boundary conditions
in a second asymptotic regime: How do you propose to
choose them!?

\noindent
{\bf Sagredo:}
 What you say sounds correct. Of course
$(X(s),\Theta(s))$ are not collective
coordinates. Nevertheless, for small
derivatives $\p_s X$ they do describe
low-energy excitations of the membrane.
There are, after all, several scales in
our problem: that of the 4D effective theory,
$\ell_{4D}$, that of the Kaluza-Klein
compactification $\ell_{KK}$, in addition
to the 11D Planck scale $\ell$. In the
domain $\ell\ll\ell_{KK} \ll \ell_{4D}$
the data of the  UV degrees
of freedom needed to make sense of the supergravity solution
are summarized at long distances by the
embedding coordinates $(X(s),\Theta(s))$. If we
only consider the modes of the 11D gravitino
$\Psi$ at scales $\ell_{4D}$ then there  is a
clear distinction between $\Psi\vert_{\Sigma}$ and
$\Theta(s)$. This coupling of degrees of freedom
from different scales is summarized by the
BST supermembrane action.

\noindent
{\bf Salviati:} Allora, non \'c\`e niente  altro da fare.
What other formalism could we use  besides that
advocated by BBS in \bbs ?
 So,  you suggest we follow
the Copenhagen interpretation?

\noindent
{\bf Sagredo:} Which is?

\noindent
{\bf Salviati:}
Shut up and calculate!

\noindent
{\bf Sagredo:} Si! Andiamo!

\subsec{The procedure in this paper}

Unfortunately, the authors of this paper
are just as confused as the Tuscan twosome.
The rules for computing brane-instanton effects
have not been clearly formulated. Indeed, upon
reflection one finds many unanswered
and vexing conceptual issues.
These issues become important
when one attempts to compute sub-leading
effects in an instanton sector.

In this note we take a practical
approach to the problem and follow the procedure
implicitly followed in the original paper of
Becker, Becker, and Strominger \bbs.
As in, for example, the  calculations of
\refs{\ahw, \ads}, we will extract
the effective superpotential by computing
instanton-induced  fermion bilinears
and then comparing to the fermion bilinears
in the low energy effective supergravity action.
The latter are given by \refs{\cfgv, \wb}
\eqn\fermbils{
\eqalign{
e^{K/2}   W   \bar \psi_{\mu  } \gamma^{\mu\nu} \bar \psi_{  \nu}
& + {i \over  4} e^{K/2}   (\CD_i W)
g^{i \bar j} (\p_j \tau_{IJ})^* \bar \lambda^I \bar\lambda^J  \cr
- e^{K/2}   \bigl( \CD_i \CD_j W\bigr)   \chi^i \chi^j
&
+ e^{K/2}
(\CD_i W) \bar\psi_{\mu}\gamma^\mu \chi^i + {\rm cplx conj.} \cr}
}
where $\CD_i$ is the metric covariant derivative for
both $\CM$ and $\CL$.

In an instanton sector an $M2$ brane wraps a
3-cycle $\Sigma \subset M_{11}$. The low
energy effective theory of the membrane is
a three-dimensional field theory with degrees of freedom
described by a super-embedding:
$\bbz: \Sigma^{3 \vert 0 } \rightarrow M^{11 \vert 32} $.
Here $\Sigma^{3 \vert 0 } $ is the
oriented brane world-volume,
and $M^{11 \vert 32}$ is eleven-dimensional space-time, thought
of as super-manifolds.  The target space is
equipped with an on-shell
supergravity background.

Following \bbs\ we formulate correlators in
an   instanton sector as follows. If we
wish to compute the 2-point function of
generic four-dimensional spacetime fermions $F_1,F_2$ at
positions $x_1,x_2$  then we must
compute a path integral within a path integral:
\eqn\indbilen{
\eqalign{
\bigl \langle F_1(x_1) F_2(x_2) \bigr \rangle_{\Sigma} &
:=
\int [D g_{\mu\nu}(x) D \psi_\mu(x) D z^i D \chi^i D\lambda(x) D A(x)]
e^{- S_{\rm 4D sugra} }\cdot \cr
&\qquad\qquad\qquad
 \cdot F_1(x_1) F_2(x_2)\cdot \int [D \bbz(s)  ] e^{- S_{M2}[\bbz(s); g
,C,\Psi] } .  \cr}
}
The first path integral is that of the effective
four-dimensional supergravity described above. The second
path integral is that of
 a three-dimensional field theory on the
$M2$-brane,  described below. The
$M2$ brane couples to  the background fields $g ,
C, \Psi$
of 11D  supergravity, and we substitute the
Kaluza-Klein ansatz.
While the above procedure surely receives many
corrections when computing generic correlation
functions, it might well be exact for the special
correlators that determine the superpotential.
It would be highly desirable to have a firmer
foundation for brane-instanton computations in
M-theory. Nevertheless, we maintain our
practical attitude and continue.

As in \bbs\ the amplitude \indbilen\ is
evaluated by contracting $F_1,F_2$ with the
coupling of $\Psi$ to the brane through the
``gravitino vertex operator''
$\sim \int_{\Sigma}   \bar \Psi V$.
The  contractions of $F_1,F_2$ with
$\Psi$ are  carried out by first substituting the
Kaluza-Klein reduction
\fermkk. The fermion  propagators are then
  amputated, as in \ads. We are then left
with the computation of
the two-point function
$\langle V(s_1) V(s_2) \rangle $ of the
``gravitino vertex operator'' in the three-dimensional  membrane
theory.
\foot{Actually, there are also $\Psi^2$ interactions
in the expansion of $S_{M2}$ in powers of
$\Psi$. These in fact do contribute to fermion bilinears
and are important in sorting out contact terms.
These (important!) subtleties will not
be discussed in the present paper.  }

In order to contribute to the
superpotential the membrane instanton must
leave at least two fermion zero-modes unbroken.
In the following we will evaluate the contribution
in the case when there are exactly two fermion
zero-modes. The possible contribution of
configurations with more fermion zero-modes
is left for future work.
In order to examine the fermion zero-modes,
and compute determinants  we need to understand
in detail the expansion of the membrane Lagrangian
in curved spaces. We describe this in the next section.

\newsec{Effective actions for $M_2$ branes in
curved spaces}

As above, we regard
an $M2$-brane as a dynamical object
described by a super-embedding
$\bbz: \Sigma^{3 \vert 0 } \rightarrow M^{11 \vert 32} $.
Separating bosonic and fermionic coordinates we have
\eqn\embed{
\bbz^{\bbm}(s) =  \bigl( X^M(s), \Theta^\mu(s) \bigr)
}
where
$\Theta(s)\in \Gamma \bigl[ \CS(TM_{11})
\vert_{\Sigma}\bigr] $ is a
section of a pulled-back spinor bundle.
These degrees of freedom are governed by
 the super-membrane action of
Bergshoeff, Sezgin, and Townsend \bst.
In    Euclidean  signature the action enters
the path integral through
\eqn\culddbf{
\exp\Biggl\{ -   \int_{\Sigma}  d^3 s
\Biggl[\ell^{-3} \sqrt{\det_{ij} g_{ij} }  - {i \over  3!}
 \epsilon^{ijk} \p_i \bbz^{\bbm}  \p_j \bbz^{\bbn}  \p_k \bbz^{\bbp }
\bbc_{\bbp\bbn\bbm }\bigl(X(s), \Theta(s) \bigr)  \Biggr]
\Biggr\}
}
where
\eqn\defpi{
\eqalign{
\Pi_{i}^{~~\bba} & = \p_{i} \bbz^{\bbm} \bbe_{\bbm}^{~~\bba} \cr
g_{ij}& = \Pi_{i}^{~~A} \Pi_{j}^{~~B} \eta_{AB} \cr}
}
The sign of the WZ term depends on the orientation of
$\Sigma$.

The action \culddbf\ has bosonic and fermionic
gauge invariances, $\diff(\Sigma) \oplus \diff_\kappa $,
where $\diff_\kappa $ refers to ``$\kappa$-supersymmetry''
\bst.
The  induced metric for any embedding of $\Sigma$ gives an
orthogonal decomposition
$TM_{11}\vert_\Sigma  = T\Sigma \perp \CN$
in terms of tangent and normal bundles.
This leads to a
 reduction of the structure group
$Spin(1,10)   \supset Spin(1,2)_{\parallel } \times Spin(8)_\perp$
under which
${\bf 32_{\IR} }   = ({\bf 2}; {\bf 8^-}) \oplus ({\bf 2}; {\bf 8^+})$.
Therefore the spinor bundles can be reduced as
\eqn\redsp{
\CS(TM_{11})\vert_{\Sigma}   = \CS(T\Sigma) \otimes \CS^-(\CN) \oplus
\CS(T\Sigma) \otimes \CS^+(\CN).}
In this language, the meaning of
$\kappa$-supersymmetry is   that
the physical degrees of freedom are in
$ \CS(T\Sigma) \otimes \CS^-(\CN) $.
Hence, after fixing static gauge   the
physical degrees of freedom are given by
\eqn\physmult{
\eqalign{
y & \in \Gamma[ \CN \rightarrow \Sigma]\cr
\Theta & \in \Gamma[\CS(T\Sigma) \otimes \CS^-(\CN)] \cr}
}
where $\CN$ is the normal bundle to $\Sigma$ in
the full eleven-dimensional space-time, and $S^-(\CN)$ is the negative
chirality   spinor bundle associated to $\CN$.

In order to describe the action for the physical
degrees of freedom in a   way useful for our
purposes we need to
expand the action in powers of $y(s), \Theta(s)$.
Let us first focus on the purely bosonic action.
Choose a local coordinate system $X^M = (x^{m'},y^{m''})$ so
that $\Sigma$ is described by the
equation $y^{m''}=0$. In general we will
denote tangent indices with a prime and normal
indices with a double-prime. We choose coordinates
so that the metric tensor on $(TM_{11} )\vert_{\Sigma} $
is of the form
\eqn\metric{
g\vert_{\Sigma}  = \pmatrix{ h_{m' n'}(x) & 0 \cr
0 & h_{m'' n''}(x) \cr}.
}
Expanding the induced area we obtain the
standard result that the area is stationary
if $\Sigma$ is a minimal sub-manifold, i.e.,
if the trace of the second fundamental
form is zero. Choosing static gauge
$x^{m'}(s) = s^i \delta_i^{~~m'}$
we find  that the expansion around a
minimal sub-manifold is:
\eqn\epdbi{
\eqalign{
\int_\Sigma d^3 s \sqrt{\det[\p_{m'} X^M \p_{n'} X^N g_{MN}(s,y(s)) ] }
& = \int_\Sigma ds \sqrt{\det h_{m'n'}(s)}
 \cr
+ \int_\Sigma ds \sqrt{\det h_{m'n'}(s)}
\biggl(   \half (D y)^2 - \biggl[ \half R^{m'}_{~~ k'' m' l''}
&
+ {1 \over  8}
Q^{m'n'}_{~~~~~ k''} Q_{m'n' l''} \biggr] y^{k''} y^{l''} + \CO(y^3)
\biggr) \cr}
}
where $D$ is the induced connection on the
normal bundle,  $Q_{m'n' l''}$ is the second
fundamental form, and $R^{m'}_{~~ k'' m' l''}$ is the ambient curvature
tensor restricted to the brane.

To obtain the action for the fermions we
expand the on-shell supergravity background in
powers of the super-coordinate $\Theta$.
The torsion constraints for
11D on shell super-space were found in
\elevss.
The frame to order $\Theta$ including the gravitino but
putting $\bbg_{ABCD}\vert= G_{ABCD}=0$ is:
\eqn\frameone{
\eqalign{
\bbe_{\bbm}^{~~\bba }   & = \pmatrix{ e_M^{~~ A} - i \bar\Psi_M \Gamma^A \Theta
& \half \Psi_M^{~ \alpha} + {1 \over  4} \omega_M^{CD}
(\Gamma_{CD})^\alpha_{~\nu} \Theta^\nu \cr
-i \Gamma^A_{\mu\nu}\Theta^\nu & \delta_\mu^{~ \alpha} \cr}
+ \CO(\Theta^2) \cr}
}
where $\omega_M^{CD}$ is the Riemannian
spin connection.
In addition, we will need
\foot{
After working  out these superspace
expansions we learned of
the paper \dpp\ which also works out the frame
to order $\Theta^2$. The conventions are different
but up to numerical coefficients our expressions
agree, at least
up to  order $\CO(\Theta^2  \Psi)$. See also
\shibusa.  }

\eqn\secord{
{\bbe}_M^{~~A}  = e_M^{~~A} -{i \over  4}
\bar\Theta  \Gamma^{ACD} \Theta  \omega_{M,CD} + \CO(\Theta^3,\Psi)
}
\eqn\ceethetexp{
\eqalign{
\bbc_{MNP}& = C_{MNP}(x)+ i \biggl(
\bar\Theta \Gamma_{MN} \Psi_P +
\bar\Theta \Gamma_{PM} \Psi_N+
\bar\Theta \Gamma_{NP} \Psi_M \biggr) \cr
& + {3 i \over  4}
 \bar\Theta \Gamma_{[MN} \Gamma^{CD} \Theta \omega_{P],CD} + \CO(\Theta^3)
\cr
\bbc_{MN\rho}& = - i (\Gamma_{MN}\Theta)_\rho +
\CO(\Theta^3) \cr
\bbc_{M\nu\rho}& =  \CO(\Theta^2) \cr
\bbc_{\mu\nu\rho}& =  \CO(\Theta^3) \cr}
}

Using the above results on the frame we find:
\eqn\pipiagain{
\eqalign{
\Pi_i^{~ A} & = e_M^{~A} \p_i X^M  - i \bar \Theta \Gamma^A
\bigl( D_i \Theta  +   \p_i X^M \Psi_M \bigr)
+\CO(\Theta^4, \Theta^3 \Psi) \cr
\Pi_i^{~ \alpha} & = (D_i \Theta^\mu )\delta_\mu^\alpha + \half
\p_i X^M \Psi_M^\alpha +
\CO(\Theta^3, \Theta^2 \Psi ) \cr}
}
where $D_i \Theta$ is the pullback of the spin connection
from the ambient space:
\eqn\pullback{
(D_i \Theta)^\mu :=
 \p_i \Theta^\mu +  \p_i X^M \omega_M^{CD} {1 \over
4}(\Gamma_{CD})^\mu_{~\nu} \Theta^\nu .
}

We now consider the expansion of
\culddbf\ in fluctuations around $y=\Theta=0$.
Treating the brane as an elementary
object   the  leading order term is
simply
\eqn\leadact{
\exp[ i \int_{\Sigma}( C + i \ell^{-3}\vol(h)    ) ].
}

Expanding around a minimal sub-manifold
and including the fermions we find the
   quadratic   action
\eqn\fullquad{
\eqalign{
\ell^{-3} \int_\Sigma d s \biggl[
\sqrt{\det h} & \biggl( h^{ij} (D_{i} y^{m''})
(D_{j} y^{n''}) h_{m'' n''} - y^{m''}\CU_{m'' n''} y^{n''}
 \biggr)\cr
 \qquad \qquad - i \sqrt{\det h} h^{ij} e_{i}^{~~a'} (\bar\Theta \Gamma^{a'}
D_{j} \Theta)    &
+
{1 \over  2} \epsilon^{ijk} \bar\Theta \Gamma_{ij} D_k \Theta
\biggr] \cr}
}
where $\Gamma_i := \p_i X^M \Gamma_M$,
and the covariant derivatives are those
determined by the above bundles.
(There are terms
 in the pulled-back spinor connection
involving the second fundamental form.
These can be shown to
vanish using the fact that
$\Sigma$ is a minimal submanifold.)
The ``mass term''
$\CU$ depends on the curvatures and second fundamental
form, as   in \epdbi. In Euclidean space the two
fermion kinetic terms are equal, to lowest order
in interactions. If we keep the gravitino in the
expansion \frameone\secord\ we find
 the coupling to the background gravitino is
$i \ell^{-3} \int_\Sigma
 d  s  \sqrt{\det h} \bar \Psi_M V^M +\CO(\Psi^2)$,
where $V^M$ is the gravitino vertex operator
\eqn\grvtnvtx{
V^M=
h^{ij} \p_i X^M \p_j X^N    \Gamma_N  \Theta
 + {1\over  2}
\epsilon^{ijk}    \p_i X^{M}  \p_j X^{N}  \p_k X^{P}
 \Gamma_{PN} \Theta.
}

Let us now consider the supersymmetries of
the action.  These will be quite useful in
sorting out topological field theories in the
next section.
The unbroken super-isometries of the
on-shell 11D background become global
supersymmetries of the M2 theory.
The super-isometries are defined by
a super-vector   $\bbk^{\bbm} = (k^M, \kappa^\mu)$
satisfying the equations:
\eqn\superisom{
\eqalign{
\delta \bbz^{\bbm} & = - \bbk^{\bbm} \cr
\delta \bbe_{\bbm}^{~~\bba} & = \bbk^{\bbn} \p_{\bbn}
\bbe_{\bbm}^{~~\bba} + \p_{\bbm}
 \bbk^{\bbn} \bbe_{\bbn}^{~~\bba} = 0 \cr}
}
Setting the background $\Psi=0$ the isometry
can be given as an expansion in powers of
$\Theta$:
\eqn\firsttem{
\eqalign{
\kappa^\mu & = \epsilon^\mu + (\Delta \epsilon)^\mu \cr
& = \epsilon^\mu  - {  i \over  4}
(\Gamma_{CD} \Theta)^\mu (\bar \Theta \Gamma^M \epsilon) \omega_M^{CD} +
\CO(\Theta^4\epsilon) \cr
k^N
& = i \bar \Theta \Gamma^N \epsilon + (\Delta k)^N \cr
& = i \bar \Theta \Gamma^N \epsilon + \CO(\Theta^3 \epsilon) \cr}
}
where $\epsilon$ is a covariantly constant spinor for the
bosonic $(\Theta=0)$ background. Thus, including the
gauge transformations, the action is invariant
under the following transformations:
\eqn\unbrokiip{
\eqalign{
\delta \Theta(s) & = \epsilon + (\Delta \epsilon) + \delta_\kappa \Theta +
v^i(s) \p_i \Theta  + \CO(\Theta^4 \kappa) \cr
\delta X^M(s) &  = i\bar \Theta \Gamma^M \epsilon + (\Delta k^M)- i \bar\Theta
\Gamma^M \delta_\kappa \Theta + v^i(s) \p_i X^M + \CO(\Theta^3 \epsilon) \cr
\delta_\kappa \Theta := &  (1 + \Gamma_{\parallel}[X,\Theta]) \kappa(s)\cr}
}
where $v^i(s)$ is a diffeomorphism of $\Sigma$,
  $\kappa(s)$ is an arbitrary
spinor and
\eqn\projpar{
\Gamma_{\parallel}[X, \Theta]:= {1 \over  3!}
{1 \over  \sqrt{\vert \det g\vert} }  \epsilon^{ijk} \Pi_i^{~ A}
\Pi_j^{~ B}\Pi_k^{~C} \Gamma_{ABC}
}
is the induced Clifford volume element.
One easily shows that
$\Gamma_{\parallel}[X, \Theta]^2 = - {\det g \over  \vert \det g \vert}$,
so, working in Minkowskian
signature $\half(1 + \Gamma_\parallel)$ is a
projector. To see  how these transformations act on
physical degrees of freedom we fix $\diff(\Sigma)$
symmetry by choosing static gauge. Then, to fix $\kappa$
supersymmetry we choose a decomposition
of spinors under
$Spin(1,10)   \supset Spin(1,2) \times Spin(8)$
such that ${\bf 32}  = (2,8^-) \oplus (2,8^+) $.
We choose a representation of the Clifford
algebra, and spinor conventions as in
appendix A. In this representation the
spinor degrees of freedom may be written as:
\eqn\spindof{
\Theta   = \pmatrix{ \Theta_1^{Aa} \cr \Theta_2^{A \dot a} \cr}.
}
For the   configuration $y=\Theta=0$
the induced Clifford volume is
$$
\Gamma_{\parallel}[y=0, \Theta=0] = 1_2 \otimes \pmatrix{ - 1_8 & 0 \cr 0 &
1_8\cr}
$$
so we fix the gauge by taking:
\eqn\gaugefixk{
\Theta = \pmatrix{ \Theta^{Aa}(s) \cr 0 \cr}.
}

In order to   get something
useful from
\unbrokiip\ we introduce the low energy expansion
which is an expansion in degrees defined by:
$[s^i] = -1 , [y^{m''}]  = 0 , [\p_i y^{m''} ]   = 1 ,
[\Theta_1]   = 1/2 $.
(Note that this differs from the energy expansion
in bulk supergravity, where $[\Theta]$ has weight
$-1/2$.)
The way in which we assign degrees
to the geometrical objects
$[e_M^{~A} ] , [\omega_M^{AB}], [R_{MNPQ}]$
requires discussion.
In general we must assign degree one to
${\p \over  \p s^i}$. If some set of
normal directions has a direct
product structure then we can consistently
assign degree zero to $[e],[\omega], [R]$ in those
normal directions. However, if there are off-diagonal
terms in the metric, e.g. second fundamental
forms or connections on the normal bundle,
then we must assign   degree
one to $  [\omega_M^{AB}], [R_{MNPQ}]$.

We now decompose the covariantly constant
spinor in terms of $Spin(8)$ chirality as in \spindof:
\eqn\spindof{
\epsilon
= \pmatrix{ \epsilon_1^{Aa} \cr \epsilon_2^{A \dot a} \cr}
}
The spinors $\epsilon_1$ and $\epsilon_2$
 generate a broken and an unbroken
supersymmetry in the membrane theory, respectively.
We consider the supersymmetry parameter to
have degree $+1/2$. Then, the degree
expansion of   the broken $\delta_1$, and unbroken $\delta_2$
supersymmetries  takes the form
\eqn\expndsus{
\eqalign{
\delta_1 & = [\delta_1]^{0} +   [\delta_1]^{+2}+ \cdots
\cr
\delta_2 & =   [\delta_2]^{+1 } + [\delta_2]^{+2}+ \cdots \cr}
}
where the superscript denotes the change in degree.
Explicitly, the broken supersymmetry is given by
\eqn\brokensi{
\eqalign{
\delta_1 y^{m''} & = 0 \cr
\delta_1 \Theta_1 & = \epsilon_1 + (\Delta \epsilon_1)_1 \cr
(\Delta \epsilon_1)_1& = - { i \over  4} (\bar \Theta_1 \tau^{a'} \epsilon_1)
(\omega_{a'}^{c'd'} \tau_{c'd'} \Theta_1 +
\omega_{a'}^{c''d''} \gamma_{c''} \tilde \gamma_{d''}  \Theta_1)\cr}
}
while the unbroken supersymmetry is
\eqn\brokensii{
\eqalign{
\delta_2 y^{m''} & = -i(\bar \Theta_1 \gamma^{a''} \epsilon_2)
e_{a''}^{~~m''}(s,y(s) )  \cr
\delta_2 \Theta_1 & = -[B]^1 \epsilon_2 + (\Delta \epsilon_2)_1 \cr
(\Delta \epsilon_2)_1& = { i \over  4} (\bar \Theta_1 \gamma^{a''}  \epsilon_2)
(\omega_{a''}^{c'd'} \tau_{c'd'} \Theta_1 +
\omega_{a''}^{c''d''} \gamma_{c''} \tilde \gamma_{d''}  \Theta_1)\cr
[B]^1 & = - \bigl( \p_i y^{m''} e_{m''}^{~~a''} +
{i \over  2} (\bar \Theta_1 \tau^{c'} \gamma^{a''} \tilde \gamma^{d''}
\Theta_1) (\omega_i)_{c'd''} \bigr) \tau^i \gamma_{a''} \cr}
}

 The unbroken transformations are closely
related to the transformations of a supersymmetric
3d sigma model. In particular note the terms:
\eqn\brokensiii{
\eqalign{
\delta_2 y^{m''} & = -i(\bar \Theta_1 \gamma^{a''} \epsilon_2)
e_{a''}^{~~m''}(s,y(s) )  \cr
\delta_2 \Theta_1 & = \bigl( \p_i y^{m''} e_{m''}^{~~a''} \bigr) \tau^i
\gamma_{a''} \epsilon_2 -  { 1 \over  4}
\omega_{m''}^{c''d''} \gamma_{c''} \tilde \gamma_{d''}  \Theta_1
(\delta_2 y^{m''}) \cr}
}
The extra terms in
\brokensii\ are related to second fundamental
forms and connections on the normal bundle.
They do not occur in the usual treatment
of supersymmetric sigma models because
the metric on the world-volume and the metric in
target space is a product metric in the standard
sigma model.

\newsec{Membranes in a $G_2$ manifold}

We now consider  manifolds of the
form $M_4 \times X$ where $M_4$ is
hyperk\"ahler and $X$ has $G_2$
holonomy. In order to induce a superpotential
the brane instanton should leave
at least 2
unbroken supersymmetries.
A cycle $\Sigma \subset X$ such that $X$ has a covariantly
constant spinor $\epsilon$ with $(1 +
\Gamma_{\parallel}[X,\Theta=0])\epsilon=0$
is   a
``supersymmetric cycle''
\bbs. A key result of \bbs\bbmo\ is that such cycles are the
``calibrated sub-manifolds'' of  Harvey and Lawson \hl.
Therefore,   we specialize to the
case of an associative 3-fold $\Sigma$ in
$X$. Under these circumstances the
formulae of the previous section simplify
considerably.

We begin this simplification by
noting that in the above circumstances
there is a reduction of
  the normal bundle structure group
\eqn\rdnrml{
Spin(8)_{\perp }
\supset
Spin(4)_{\hat 1, \hat 2,  \hat 3, \hat 4}\times
Spin(4)_{\check 1, \check 2,  \check 3, \check 4  }
}
Under this reduction the general
${\rm Spin}(8)$ spinor can be written as
\eqn\genspin{
\psi = ( (\psi_{--})_{\alpha}^{~~ Y},
(\psi_{++})_{\dot \alpha}^{~~ \dot Y};
(\psi_{+-})^{\dot \alpha  Y},
(\psi_{-+})^{\alpha \dot  Y})
}
where the $+,-$ refers to the chiralities in
the first and second ${\rm Spin}(4)$ factors.
In this notation,
after we fix $\kappa$-symmetry, and restore the
$Spin(1,2) $ spinor index $A$
for the structure group of the spinor bundle
$\CS(T\Sigma)$, we get the gauge
fixed spinor degree of freedom:
\eqn\physpin{
\Theta   = (( \Theta_{--})_{ \alpha}^{~AY} ,
( \Theta_{++})^{\dot Y}_{~~ \dot \alpha A} ; 0,0 )
}

Now we must review a few consequences of
$G_2$ holonomy, in down-to-earth terms.
On a $G_2$ manifold the $Spin(7)$ structure group
of the tangent bundle is reduced to $G_2$.
We choose an identification
\eqn\geetoo{
(\Phi, T X) \cong (\varphi_0, \IR^7 = \Im  \bbo )
}
where
in a local orthonormal frame $e^{1,2,\dots, 7}$ for
$T^* X$ we have
\eqn\locphi{
\varphi_0 = e^{123} + e^{145}   + e^{176} + e^{246}
+ e^{257} + e^{347} + e^{653}  .
}
We may identify the spin representation
of ${\rm spin}(7)$
with the octonions  $\bbo$. These, in turn,
are identified with pairs of quaternions
\michlaw.  Along an
associative 3-fold we have a covariantly
constant identification of
$T_s \Sigma \subset T_sX$, $s\in \Sigma$,
 with
$(\im \bbh, 0 ) \subset \im \bbo$.
The structure group of $T \Sigma$
has spin indices $A,B,\dots, = 1, 2$
while that of the normal bundle
has $SU(2) \times SU(2)$ spin indices
$Y, \dot Y$, respectively. The corresponding
spin bundles are denoted $E_\pm$.
The spin
connection on $X$, restricted to $\Sigma$,
 may be written as
\eqn\spincn{
  \omega
= \pmatrix{ (\omega_{\parallel})^{a' b'} & \omega^{a' b''
}\cr
-\omega^{b' a'' } & (\omega_{\perp})^{a'' b'' }\cr} \qquad }
where $a',b'=1,2,3$ are tangent frame indices and
$a'',b''=4,5,6,7$ are normal bundle orthonormal  frame indices.
This corresponds to the decomposition of the
adjoint of $so(7)$ under the $so(3)_\parallel \oplus so(4)_\perp$ subgroups.
The adjoint of $so(4)$ can be further decomposed in
terms of its self-dual and antiselfdual parts with
spinor indices $(\omega_+)^{\dot Y}_{~~ \dot Y'}$
and $ (\omega_-)^{  Y}_{~~  Y'}$.
Similarly, the off-diagonal connection  has
spinor indices $(\omega)^{\dot Y Y A B} $
symmetric in $AB$.
Writing a spinor of $spin(7)$ as
\eqn\spindn{
\psi   =(  (\psi^-)^A_{~~Y} ,  (\psi^+)^{\dot Y}_{~~A})
}
(regarded as a pair of quaternions)
the covariantly constant spinor is
$\vartheta= (\delta^A_{~~Y} , 0)$. Writing out
the covariant constancy condition
$\nabla_m \vartheta=0$ we find
 the spin connection satisfies
\eqn\idconn{
(\omega_{\parallel })^{A}_{~~Y} - (\omega_{-})^{A}_{~~Y} = 0
}
and
\eqn\conoth{
 \omega^{\dot Y Y} _{~~~~ Y A} = 0,
}
that is, $(\omega)^{\dot Y Y A B} $ is totally
symmetric in the undotted indices.
The connection $\omega_+$ is unconstrained.
These equations describe the decomposition of
the adjoint ${\bf  14}$ of $ {\bf g_2} \subset
{\bf \rm spin(7)}$ under
the $so(3) \oplus so(4)$ subgroup
of ${\bf \rm spin(7)}$.

The condition \idconn\ is very important and
is one concrete way to understand the connection
of brane actions to topological field theory,
(This was predicted in \bsv\ based on
R-symmetry considerations.) The identification of
connections for the tangent bundle $\omega_\parallel$
and  the normal bundle, $\omega_-$
(with ``R-symmetry'' structure
group)  is a standard approach to
formulating the
procedure of ``topological twisting''
\tqft. (See \refs{\cohoft, \bbrt, \cmrtft} for   reviews.)

 It follows from \idconn\
that the $G_2$ structure allows us to identify the
spin groups transforming the indices $A$ and $Y$.
Thus we may may trade in the fermions
\eqn\defrw{
( \Theta_{--})_{ \alpha}^{AY} = \eta_\alpha \varepsilon^{AY} + \chi_{j \alpha}
(i \tau^j)^{AY}
}
for zero-forms $\eta$ and 1-forms $\chi$ on $\Sigma$,
as is familiar in topological field theory.
(These still carry an index $\alpha$ since they are
spinors on $M_4$.)
Moreover, again because of \idconn,  the normal
bundle directions for motion of $\Sigma$
within $X$, $y^{\check m}$, can be written as
a bispinor
$y^{A \dot Y}$, that is, as sections of
$\CS(T\Sigma) \otimes E_+$.  Thus, the physical degrees of
freedom of the $M2$ brane naturally
split into  two multiplets reflecting the
reduction of the structure group
\rdnrml.
The first multiplet, whose bosonic degrees of
freedom describe normal motion in $M_4$
is
\eqn\rwmult{
(y^{\alpha \dot \alpha} , \eta_{\alpha} , \chi_{j\alpha})
}
We call this the ``Rozansky-Witten multiplet''
(or just the RW multiplet).
Denoting $\nu_{\dot \alpha A}^{~~\dot Y} =
 ( \Theta_{++})_{ \dot \alpha A}^{\dot Y}$
the second multiplet is
 $(y^{A \dot Y} , \nu_{\dot \alpha A}^{~~\dot Y})$.
We call this the ``McLean multiplet.''

We will now explain why these names are
appropriate.
Because of the $G_2$ holonomy, the covariantly constant
spinor on $M_4 \times X$  must have the
form:
\eqn\covcnsti{
\epsilon   = (\epsilon_1;\epsilon_2) =
(\epsilon^- _{ \alpha}  \varepsilon^{AY} , 0  ;
\epsilon_+ ^{\dot\alpha  } \varepsilon^{AY}  ,0)
}
where $\epsilon^- _{ \alpha} ,
\epsilon_+ ^{\dot\alpha  } $ are
negative and positive chirality
covariantly constant spinors on the
hyperk\"ahler manifold $M_4$, respectively.
As in the previous section, these give
linear and nonlinear supersymmetries.

Under the nonlinear supersymmetry all fields transform to zero except
$\delta \eta_{\alpha}=\epsilon^- _{ \alpha}$.
Under the linear supersymmetry the RW multiplet transforms as:
\eqn\rwlinsusy{
\eqalign{
\delta_\epsilon y^{\alpha \dot \beta} & = - 2i \eta^\alpha  \epsilon_+^{\dot
\beta}   \cr
\delta_\epsilon
(\Theta_{--})_{ \alpha}^{AY} & = - (\p_iy)_{\alpha \dot \beta} \epsilon_+^{\dot
\beta}
- (\delta_\epsilon y^{\hat m})(\omega_{\hat m})_{\alpha}^{~~\beta}
(\Theta_{--})_{ \beta}^{AY}\cr}
}
while from
\brokensii\  we find
 the McLean multiplet transforms as:
\eqn\mcleanmult{
\eqalign{
\delta y^{A\dot Y} & = i \nu^{\dot \alpha A \dot Y} \epsilon_{+\dot \alpha} \cr
\delta \nu_{\dot \alpha A \dot Y} & = - (\Dsl y)_{A\dot Y} \epsilon_{+ \dot
\alpha} . \cr}
}

The transformations  \rwlinsusy\
are just the transformations of Rozansky and Witten \rw.
\foot{Equation 2.18 in \rw\ has a
misprint.}
Moreover, from the second line of \mcleanmult\
we immediately
recover the fact that the tangent space to the
moduli of associative threefolds at $\Sigma$
is identified with the space of
zero-modes of the twisted Dirac operator.
This is Theorem 5.2 of \mclean.

The quadratic action likewise splits up as a
sum for the Rozansky-Witten and McLean
multiplets. The action for the RW multiplet
is  \rw:
\eqn\quadfluct{
\int_{\Sigma}  dy^{\alpha \dot \alpha} \wedge * dy_{\alpha \dot \alpha}  +
\epsilon_{ \alpha   \beta}
\bigl( \chi^{  \alpha} \wedge * d \vartheta^{  \beta} +
\chi^{  \alpha} \wedge   d \chi^{  \beta}\bigr)
+ R_{\alpha\beta\gamma\delta}
\eta^\alpha \chi^\beta\wedge \chi^\gamma
\wedge \chi^\delta + \cdots
}
where  $R_{\alpha\beta\gamma\delta} $
is the self-dual part of the curvature,
and we have not been careful about numerical
coefficients. We have only shown terms up
to degree 2. Although we only worked out
the quadratic action in \fullquad, invariance under
\rwlinsusy\ requires the curvature term.

Similarly, the quadratic fluctuations for
the McLean multiplet are given by:
\eqn\quadflctii{
\int  [   ( \Dsl_{E_+} y )^2 + \bar \nu \Dsl_{E_+} \nu  + \cdots ]
}
here $\Dsl_{E_+}$ is the Dirac operator  twisted
by $E_+$ (and again, we have not been
careful about numerical coefficients).
  The bosonic terms in the action correspond to
Theorem 5.3 of \mclean. It follows from
\expndsus\ that the  higher order
terms in \quadflctii\ are of degree 3.

It is interesting to note what happens when we
specialize the holonomy of $X$ further and take
$X = Z \times S^1$ where $Z$ is a Calabi-Yau
3-fold and $\Sigma \subset Z$ is a special
Lagrangian submanifold. Then the ${\rm Spin(4)}$
structure group of $\CN(\Sigma \hookrightarrow X)$
is reduced to the ${\rm Spin(3)}$ structure group
of $\CN(\Sigma \hookrightarrow Z)$. The group
$Spin(3)=SU(2)$ is embedded diagonally in
$Spin(4)$ so we have another identification of
connections $\omega_+ = \omega_-$. We can therefore
introduce a topological twisting of the McLean
multiplet
\eqn\toptwoo{
\Theta^{ \dot \alpha A\dot Y}   = \tilde \eta^{\dot \alpha }\varepsilon^{A \dot
Y} + \tilde \chi_{j }^{\dot\alpha}  (i \tau^j)^{A\dot Y} .
}
The first order deformations of $\Sigma$ in
$Z$ are now given by zeromodes $\tilde \chi \in \CH^1(\Sigma;\IR)$. In this way
we can reproduce
McLean's theorem 3-6 on deformations of special
Lagrangian submanifolds of $Z$ \mclean.

\newsec{Computation of the superpotential}

\subsec{Zeromodes and Determinants}

In this section we compute the contributions
from M2 instantons to the superpotential for
$G_2$ compactification for the choice  $M_4 = \IR^4$.
We have seen that at degree 2 the M2 theory is a
product of a theory for the RW multiplet and for the
McLean multiplet. We now describe the zeromodes
and one-loop determinants for these multiplets.

The zero-modes in the path integral for the
RW multiplet are:
(1.)  The constant scalars
$y^{\alpha \dot \alpha}(s) = y^{\alpha \dot \alpha}_0$.
These simply correspond to the position of the susy
3-cycle in $\IR^4$.
(2.) The 2 constant fermion zero-mode partners
$\eta^{  \alpha}(s) = \vartheta^{  \alpha}_0$.
(3.) Harmonic 1-form zero-modes for   $\chi$.
There are $b_1(\Sigma) $ such  linearly independent
zero-modes.

As discussed in \rw\ the path integral   is a measure on the
space of zero-modes $\Lambda^{max} (2H_0(\Sigma;\IR)) \otimes
\Lambda^{max}(H_1(\Sigma;\IR) )$. The measure is the product of the natural
measure
given by the metric with the $\zeta$-regularized
determinants. As discussed in \rw\ the path integral
takes the form
\eqn\ellmin{
{\det' L_- \over (\det' \Delta^0)^2}
}
where $\Delta^{(0)}$ is the Laplacian on
scalars and $L_-$ is an operator
appearing in Chern-Simons perturbation
theory  \jones.
The absolute value of the determinants is the
Ray-Singer torsion. The phase is more subtle
and is discussed below.

We now turn to the McLean multiplet. The
path integral gives a measure on the space
of zero-modes $[d y^{A\dot Y}_0 d \nu_0]$
and therefore gives a measure on the
moduli space of supersymmetric 3-cycles
deformable to $\Sigma$.  In this case  the
quadratic fluctuation determinant is just
the phase of the Dirac determinant
\eqn\mcldet{
{\Det' (\Dsl_{E_+}) \over  \sqrt{ \Det' \Dsl_{E_+}^\dagger \Dsl_{E_+}} }.
}
 Again this phase is subtle, and
discussed below.

Let us now specialize to $\Sigma$'s which
are rigid and topologically
rational homology spheres. Then the
measure becomes
\eqn\onelpint{
\int dy_0^{\alpha \dot \alpha} d \eta_0^\alpha
\vert H_1(\Sigma;\IZ)\vert
e^{i \alpha(\Sigma, \CN)}
\exp[ i \int_{\Sigma}( C+ i \ell^{-3}\vol(h)    ) ]
}
Here we have used the standard result
that  the Ray-Singer
torsion of a rational homology sphere is
just the order of the finite group
$ H_1(\Sigma;\IZ)$ \rw. Here
$e^{i \alpha(\Sigma, \CN)}$ is the phase from the
fermion determinants.

\subsec{The phase of the one-loop measure}

The phase of the
Dirac determinant coupled to a vector bundle
 in an odd-dimensional space-time  is
discussed in \refs{\lagew,\redlich, \agdpm, \singer,\wittga}.
One of the standard approaches to the subject is
to use a Pauli-Villars regulator. This leads to the
result
\eqn\phasedirc{
{\det' \Dsl_E \over  \vert \det' \Dsl_E\vert }
=\exp[   \mp  { i \pi \over  2} \eta(\Dsl_E) ]
}
where $\eta$ is the Atiyah-Patodi-Singer (APS)
 invariant. The sign
in the exponent depends
on the choice of sign of the mass of
a Pauli-Villars regulator.

If we use this result in
the present problem then we encounter
a problem.
The resulting expression
for the superpotential  violates
holomorphy of $W$  as a function of
$C  + i \Phi$,  because the $\eta$-invariants
such as \phasedirc\
 are nontrivial functions of the $G_2$
structure, but are not functions of $C$.
One can approach this problem by
adding local Chern-Simons terms in $\omega_\pm$
to cancel the holomorphy anomaly, but then
one runs into thorny issues related to correctly
defining the Chern-Simons terms
(which have   half-integer coefficients).
A better way to define the phase of the
one loop determinants has been
described by Witten in \wittendb\ in
the case of D1 instantons. His method
 is easily adapted to our problem.

Quite generally, when considering
membrane instantons we have
  a closed oriented three-manifold
$\Sigma$  in an oriented, spin, 11-manifold $M_{11}$
with Ricci flat metric $g$ and real rank 32
spin bundle $\CS(TM_{11})$. Restricing $\CS(TM_{11})$
  to $\Sigma$ we have the splitting \redsp.
Consider the Dirac operator $\Dsl$ on $\Sigma$
coupled to the induced connection
on $\CS(T\Sigma) \otimes \CS^-(N)$ from
the ambient  metric $g$.
The phase of the membrane determinant
involves
\eqn\membdet{
{\rm Pfaff}(\Dsl_{S^-(N)} ) \exp[i \int_\Sigma (C+i \Phi)]
}
As stressed in \fourflux\wittendb, because of the
subtle geometrical nature of the field $C$,
neither factor in \membdet\ is
separately well-defined, in general.
The fermion determinant is not well-defined
because of global anomalies. Nevertheless,
the product is well-defined. (In the
example of a $G_2$ manifold each factor
is well-defined, but not canonically so. )

Let use choose a basis of
cycles $\Sigma_i$ for $H_3(X;\IZ)$. Then
the phases   $e^{i \theta_i}$
for the expression
\membdet\ evaluated for the cycles
$\Sigma_i$ constitutes
part of the data determining
the $M$-theory background. Once this
data has been specified  the phases of
all other membrane instantons can be
expressed in terms of $e^{i \theta_i}$.
To see this consider an arbitrary
sum of cycles $\Sigma_i$ trivial in homology,
that is, such that there is
an open 4-manifold such that
$\sum [\Sigma_i] = \p M_4$. By a natural
extension of the result of \daifreed\
there should be a canonical
trivialization of the the Pfaffian line bundles
$T: \otimes_i {\rm PFAFF}(\Dsl_{S^-(N), \Sigma_i}  )
\rightarrow \IR$. Then the phase of the
product of the determinants for $\Sigma_i$ is that
of
\eqn\product{
T \biggl[ \otimes_i {\rm Pfaff}(\Dsl_{S^-(N), \Sigma_i})
\biggr]
\exp[ i \int_{M_4} G]
}
and the expression \product\
may be shown to be independent
of the choice of cobordism $M_4$.

The upshot of this discussion is that, if
we are willing to ignore the correct geometrical
status of $C$ then we can absorb the phase
$e^{i \alpha(\Sigma,\CN)}$ in \onelpint\ into
the definition of the object $\exp[i \int_\Sigma C]$.

\subsec{Computing $W$}

Let us now turn to the computation of the
fermion two-point functions in the four-dimensional
effective action. The computation of
gaugino and gravitino mass terms is
tricky because of   contact terms.
 The cleanest term to evaluate
is the part of the chiral multiplet mass term
proportional to $\p_i \p_j W$.
Computing the two-point function
$\langle \chi^{i}(x_1) \chi^j(x_2) \rangle$
for $\vert x_1 - x_2 \vert \gg \ell$ using
the instanton approximation in \indbilen\ we find
(after truncating the fermion propagators) the
mass term:
\eqn\llonelpmtii{
\eqalign{
-  \CD^{{\rm sugra}}_{\Sigma} \chi^i \chi^j v_i v_j
\biggl\vert { \det' L_- \over  (\det' \Delta^{(0)} )^2 } \biggr \vert
&
\exp[ i \int_{\Sigma}(  C + i \ell^{-3}\vol(h) ) ] \cr}
}
Here $\CD^{{\rm sugra}}_{\Sigma} $ stand
for the 11D bulk
supergravity determinants of $g,C, \Psi$
are are discussed further below. We also have:
\eqn\newdih{
v_i  = 4  \int_{\Sigma}  \omega_i^{(3)}
}
and we have taken the harmonic
three-forms $ \omega_i^{(3)}$
to be in the ${\bf 27}$ representation of $G_2$,
 again to avoid subtleties with contact
terms.

This computation shows that  a rigid rational
homology 3-sphere gives a contribution to
the superpotential of:
\eqn\ansforw{
\Delta W \propto
\CD^{{\rm sugra}}_{\Sigma}  \vert H_1(\Sigma;\IZ)\vert
\exp[ i \int_{\Sigma}(  C  + i  \ell^{-3} \Phi) ]}
where we have used the fact that
$\Phi\vert_{\Sigma} =  \vol(h)$ for
an associative 3-cycle. We only
write $\propto$ because we have
not been careful about the overall
normalization of the K\"ahler potential.
(This will involve a purely numerical factor
with  powers of $2$ and $\pi$.)
 Note that, since we are working
in a Green-Schwarz-like formalism we
do {\it not} sum over spin structures on $\Sigma$.
Rather, the spin structure on $\Sigma$ is induced
by that of the ambient supergravity background.

When we consider the supergravity determinants
$\CD^{{\rm sugra}}_{\Sigma} $  we run into
some of the conceptual problems discussed in
section three.  If  we treat the membrane
as an elementary brane in the $G_2$ manifold
$\IR^4 \times X_7$ with no account of backreaction
then we can simply take $\CD^{{\rm sugra}}_{\Sigma} $
to be a constant, independent of $\Sigma$.
Therefore,  if we treat the membrane as
an elementary membrane in a smooth
supergravity background, then we can choose an
additive constant in the Kahler potential so
that rigid membranes contribute to the
superpotential
\eqn\ansforwp{
\Delta W =
   \vert H_1(\Sigma;\IZ)\vert
 \exp[ i \int_{\Sigma}(  C  + i  \ell^{-3} \Phi) ] .
}

\subsec{Further development}

An important check on \ansforw\ will be the
computation of other terms in \fermbils.
These other terms will be more difficult to
compute because of contact terms. (For
example, the gravitino mass term is entirely
due to contact terms.)

The above discussion should also be extended to
the case of supersymmetric 3-cycles with
$b_1(\Sigma)>0$ and to non-rigid supersymmetric 3-cycles.
Since wrapped membrane instantons can sometimes
be
related to worldsheet instantons it is clear that
in general nonisolated membranes can contribute
to nonperturbative effects.
It is quite likely  that higher order interactions
from the expansion of the DBI action can
come down from the exponential and soak up
the extra zero-modes.  We expect that this will
occur already at degree three in the low energy
expansion.

Finally, one needs to
address the issue of a multiple-cover formula.
We give an example below which shows
that in general there must be contributions
from such multiple covers.
Based on the known multiple-cover
formula for the  contribution of worldsheet
instantons one may guess
that a $k$-fold covering membrane enters
as $m(k, b_1(\Sigma)) e^{-k \vol(\Sigma)} $,
where $m$ is a universal function,
but we do not have much evidence for this.

\newsec{Examples with rigid supersymmetric 3-cycles }

In this section we give an example of a class of
compactifications in which one can show that
the effects we have discussed above really exist.

\subsec{Barely $G_2$ manifolds and their moduli}

 We  now
 focus on a very special class of $G_2$ manifolds
which we call {\it barely $G_2$ manifolds}.
\foot{These manifolds are discussed briefly in
\joycei.} These are,
by definition, $G_2$ manifolds of  the form:
\eqn\exsev{
X = (Z \times S^1)/\IZ_2
}
Here $Z$ is a Calabi-Yau 3-fold.
$\IZ_2$ acts by $(\sigma_Z,-1)$. $\sigma_Z$ is
a {\it real structure}, by which we mean an
anti-holomorphic involutive
isometry of the Calabi-Yau  metric  taking
\eqn\condsigzee{
\eqalign{
\sigma_Z^*(J) & = - J \cr
\sigma_Z^*(\Omega) & = +\bar  \Omega . \cr}
}
We will take $\sigma_Z$ to act without fixed
points so $X$ is smooth.  For generic holonomy
of $Z$ the holonomy is merely $SU(3) \sdtimes \IZ_2$.
Nevertheless, there is only one covariantly
constant spinor, so for many physical
purposes $X$ is a genuine
$G_2$ manifold. Examples of smooth
barely $G_2$ manifolds   may be easily constructed
 from  complete intersection
Calabi-Yau manifolds in weighted projective
spaces defined by polynomials with
real coefficients. The case where
$\sigma_Z$ does have fixed points
can presumably be smoothed, as discussed in
\joycei. In such manifolds, some of the
following considerations will still apply.

The cohomology of barely $G_2$
manifolds is easily computed in terms of that
of $Z$ :
\eqn\hthre{
H^3(X ;\IR) = H^2(Z;\IR)^- \oplus H^3(Z; \IR)^+
}
where the superscripts denote the $\pm$ eigenspaces
under the action of $\sigma_Z^* $.
Now  suppose that  $Z$ is a Calabi-Yau obtained from a
hypersurface of a toric variety.
Then $H^2(Z;\IR)^-$ can be interpreted as
the space of  K\"ahler deformations
   inherited from the ambient quasi-projective
spaces, while $H^3(Z; \IR)^+$ may be
identified with the complex structure deformations
preserving the real structure.  Thus, we expect that
 the Calabi-Yau's with
anti-holomorphic involution $\sigma_Z^*$ form a
real sub-manifold of $\CM_{2,1} \times \CM_{1,1}$,
of {\it real} dimension $h^{2,1}(Z) +   h^{1,1}(Z)$,
\eqn\realslc{
\CM_{2,1}^{\IR}  \times \CM_{1,1}^{\IR}
\subset \CM_{2,1} \times \CM_{1,1}.
}
Indeed we expect
\realslc\ to be   a natural Lagrangian sub-manifold of
the full CY moduli  $\CM_{2,1} \times \CM_{1,1}$
(  very much in the spirit of the
discussion in  \hitchin).

We can describe the
moduli space of barely $G_2$ manifolds.
For $G_2$ manifolds
the tangent to moduli space is $H^3(X;\IR)$.
We now see that   the deformation space has a natural
interpretation in terms of the K\"ahler and complex
structure deformations preserving the isometry
$\sigma_Z^*$. Note that $H^{3,+}$ has
real dimension $  h^{1,2}(Z)+ 1$.  The
extra $(+1)$ dimension
(relative to $\dim \CM_{2,1}$)
 corresponds the normalization
of the $(3,0)$ form $\Omega$. The latter is
relevant since, in a local orthonormal frame
the calibration must look like
\eqn\anothnorm{
\eqalign{
\varphi_0 = e^{123} + e^{145} & + e^{176} + e^{246}
+ e^{257} + e^{347} + e^{653} \cr
=   J\wedge  e^1 & +
{\rm Re} [\Omega] \cr}
}
and the normalization of $\Omega$ is fixed
by ${i \over  2} \Omega \wedge \bar \Omega =
{1 \over  6} J\wedge J \wedge J$.
Here we are using a real orthonormal  frame $e^{1,2,\dots, 7}$ for
$T^* X$ from which we construct an orthonormal frame for
$T^{*1,0}Z$ given by
$w^1 = e^2+ i e^3, w^2 = e^4 + i e^5, w^3= e^6-i e^7$
and $\Omega   =   w^1\wedge w^2 \wedge w^3  $.
The  extra modulus can also be identified
with the extra degree of freedom in
the radius $R$ of the  circle  cotangent to
$e^1$. In any case, the moduli space
 of
barely $G_2$ manifolds is a real line bundle
\eqn\modbg{
\matrix{ \IR_+ & \rightarrow & \CM_{G_2}  \cr
&  & \downarrow\cr
&  & \CM_{2,1}^{\IR} \times \CM_{1,1}^{\IR}  \cr}
}
A word of caution is required here.
In applications to physics we would regard
$\CM_{G_2}$ as a real submanifold of
the K\"ahler target $\CM$ for the chiral scalars.
Such a description is valid   near a large K\"ahler
and complex structure boundary, but will
receive corrections away from such boundaries
because quantum corrections
will change $K$ and could even change the
topology of the moduli space $\CM$.

\subsec{Supersymmetric cycles}

The associative 3-cycles in
barely $G_2$ manifolds are also  easily described.
These fall into two classes analogous to
the divisors of types b,a (respectively) in sec. 3 of
\wfv.  The first class,
which we refer to as
{\it ``holomorphic 3-cycles''} are of the form
\eqn\cyclone{
[\Sigma^{\rm hol} ] \equiv ([\Sigma_2^-] \times S^1)/\IZ_2
}
where $[\Sigma_2^-] $ is a
holomorphic cycle in $Z$ mapped to
$-[\Sigma_2^-] $ by $\sigma_Z$. That is,
$\sigma_Z$ preserves $\Sigma^{\rm hol}$
setwise and  is orientation reversing.
These cycles will be rigid if
  $[\Sigma_2^-] $ is a rigid rational
curve in $Z$. If $\Sigma_2^-$ is a rational
curve then $\Sigma^{\rm hol}$ has the
topology  of the nontrivial
circle bundle over $\IR P^2$ and is hence
a rational homology sphere.
The second class, which we refer to as
{\it ``Lagrangian 3-cycles,''}
are of the form:
\eqn\cycltwo{
\Sigma^{\rm lag} = \Sigma^+  /\IZ_2
}
where $[\Sigma^+] $ is a cycle mapped to
$+[\Sigma^+] $ by $\sigma_Z$. Examples of
such cycles are provided by special Lagrangian
spheres in $Z$ on which $\sigma_Z$ acts
nontrivially.   In appendix C we give an example
of such a rigid Lagrangian 3-cycle in a barely
$G_2$ manifold.

\subsec{$M$-theory on barely $G_2$ manifolds}

We now consider the superpotential
for $M$-theory compactification on a barely $G_2$
manifold $X$.
Note that the real structure $\sigma_Z$ is
orientation reversing on the 6-cycle $Z$,
and therefore $H_6(X;\IQ ) = 0$. Thus, there are no
5-brane instantons in these compactifications.
Therefore, the only other nonperturbative
contributions (coming from wrapped branes) are
from membranes, and this gives
a sum over the two  sets of cycles described above:
\eqn\sumcyc{
W =  W_{\rm hol}(Z) + W_{\rm lag}(Z)
}
where $W_{\rm hol}(Z)$ is the sum over holomorphic
curves in $Z$ and $W_{\rm lag}(Z)$ is the sum over
susy 3-cycles in $Z$ with the above
orientation properties. Note these can be expressed
purely in terms of the data of $Z$ and its
holomorphic $(3,0)$ form.


\newsec{Superpotentials in Type IIA compactifications}

The compactifications we have
discussed are closely related to
compactifications
of IIA string theory on  $G_2$ manifolds
to three dimensions. Once again,
there will be instanton-induced
superpotentials, arising from
many new effects. In the first
place, there can now be
worldsheet instanton effects. Moreover,
there will be D-brane instanton
effects associated to $p=0,2,4,6$
branes wrapping nontrivial
classes in $H_1, H_3, H_5, H_7$,
respectively. If we make the mild
\foot{For example, for a barely
$G_2$ manifold it suffices
to take $b_1(Z)=0$ and
$H^{2,+}(Z)=0$. The latter
condition follows if all the
K\"ahler classes are induced
from an ambient projective space.}
assumption that
$H_1(X) = H_2(X) = H_5(X) =0$
then there are no
   worldsheet instanton effects,
and no $D0$ or $D4$ instanton
effects. The computation of
D2 instantons must be
equivalent to the above
results   since the M2 and D2 brane
Lagrangians can be mapped into one another
by a change of variables \townsend.
Thus, the main new unavoidable
ingredient is the effect of  $D6$
instantons.

Let us consider briefly the effects of
singly-wrapped  $D6$ instantons,
temporarily relaxing the restriction
$H^2(X)=0$ . The action for an
elementary $D6$ brane wrapped on
$X$ is
\eqn\eldsx{
e^{i   \bigl(\gamma + i \ell^{-7} \vol(X)\bigr)}
}
where $\gamma = \int C^{(7)}$ and the volume
is measured in $M$-theory units.
This follows from the Born-Infeld
action in IIA theory. It can also be obtained
directly from $M$-theory using
  the identification
 between the D6 brane of type IIA
supergravity and the Kaluza-Klein monopole
of 11D supergravity, $TN \times X$, where
$TN$ is the singly charged Taub-NUT
space \townrev.
The metric at infinity asymptotes to the
metric of $\IR^3 \times S^1 \times X$.
Putting a boundary $S^2 \times S^1 \times X$
near infinity where $S^2$ has radius $r$
we compute the $M$-theory Einstein-Hilbert
action:
\eqn\ehaction{
\eqalign{
\lim_{r \rightarrow \infty}
\Biggl[ {1 \over  (2 \pi)^2 \ell^9}
\biggl(\int_{\IR^3 \times S^1 \times X}
\sqrt{g} \CR(g) +   \int_{S^2 \times S^1 \times X} Q \biggr)
& \qquad\qquad\qquad\cr
-  {1 \over  (2 \pi)^2 \ell^9} \biggl(\int_{TN \times X}
\sqrt{g} \CR(g) +&    \int_{\p TN \times X} Q \biggr)
\Biggr]
  = \vol(X)/\ell^7\cr}
}
where in the first term we have the flat metric on
$\IR^3 \times S^1$, and $Q$ is the second fundamental
form (normalized as in $(A.1)$).

In addition a single D6 brane carries an
abelian super Maxwell  theory. When
the fieldstrength ${ 1\over  (2 \pi)}\CF $ is in
$\CH^2(X; \IZ)$ the Born-Infeld action
gives a correction to the
instanton action. Assuming the background
IIA field $C^{(5)}=0$, the correction is
\eqn\maxact{{ i \over  8 \pi^2}    \int (C + i \ell^{-3} \Phi)
\CF \wedge \CF
 - { i \over  48} \int (C + i \ell^{-3} \Phi)  \wedge p_1(TX)
}
where  we have transcribed the result to the
$M$-theory variables $C,\Phi$.
%
%
The imaginary
part comes from the standard D-brane
Chern-Simons coupling. In the second term
we have added the real part dictated by
holomorphy. Note that  from \joycei,
Lemma 1.1.2,    $\int \Phi \wedge p_1(TX) <0$,
for  $G_2$-manifolds.

Now let us consider determinants and
zeromodes.
When ${\CF \over  (2 \pi)} $  is harmonic
there are exactly two fermion
zeromodes (if $X$ has only one covariantly
constant spinor) and the instanton
can contribute to the
superpotential. The determinants for the
fields in the super Maxwell multiplet are just
\eqn\maxdets{
\det{'} \dsl \cdot {\det' \Delta^{(0)}  \over  (\det \Delta^{(1)})^{1/2}}
\cdot {1 \over
  (\det' \Delta^{(0)} )^{3/2} } = {\det' \dsl
\over  \vert \det' \dsl \vert}
}
Here  $\det \Delta^{(1)}$ is the Laplacian on
one-forms and we assume there are no
harmonic one-forms on $X$.
Equation \maxdets\ is established by
noting that
because $\CR_{ij}=0$ on $X$ we have
$\dsl^2 = \nabla^2$
where $\nabla$ is the spinor covariant derivative.
Since $S(T X) \cong \Omega^0(X) \oplus \Omega^1(X)$
on a $G_2$ manifold the spinor covariant derivative
may be identified with the ordinary one and
the above partition function reduces to a pure
phase.   As in our discussion of the phase of membrane
determinants one could use Pauli-Villars
regularization to get a phase in terms of the
$\eta$ invariant,
$\exp[\pm {i \pi \over  2} \eta(\dsl)]$,
where
$\dsl$ is the Dirac operator on $X$.
However, this then requires the addition
of counterterms to
cancel holomorphy anomalies. Again
we can instead define the determinant
to be real, up to a sign, and the sign
ambiguity cancels against that of
\eldsx, as in \wittendb.

Putting
all these remarks together we find
the contribution of the wrapped D6-brane
 is thus
essentially a $\Theta$-function:
\eqn\sevendee{
Z_1(X) =  e^{i \rho}
\sum_{\lambda \in H^2(X,\IZ)} e^{ i { \pi \over  2} \tau_{IJ} \lambda^I
\lambda^J }
}
where
\eqn\phaseanom{
e^{i \rho} = \exp \biggl[ i (\gamma + i \ell^{-7} \vol(X))
- { i \over  48} \int (C + i \ell^{-3} \Phi)  \wedge p_1(TX)
\biggr]
}

In addition to singly wrapped $D6$ instantons
there can be multiply-wrapped instantons.
In $M$-theory these will be   $A_n$ Taub-NUT
singularities with instanton action
\eqn\aennsing{
e^{i  n \bigl(\gamma + i \ell^{-7} \vol(X)\bigr)}
}
The one-loop contributions $Z_n(X)$  for
 $n>1$ could be extremely interesting and involve the
analog of Donaldson theory partition functions for the
octonionic instanton equations in seven dimensions
\refs{\octinst,\dontom, \bks,\fks,\losev}. It is worth
noting that if $Z$ has a real structure
then Hermitian Yang-Mills connections
invariant under $\sigma_Z$ provide nontrivial
examples of solutions to the {\it nonabelian}
octonionic instanton equations on the associated
barely $G_2$ manifold $X$. This gives the
first example (of which we are aware)
 where these instanton moduli spaces are
nonempty.

In addition there will
be mixed  $D2+D6$ instanton effects. These
correspond to ``pointlike instantons'' in the
abelian gauge theory, but can be more
directly analyzed by considering
M2 branes wrapping $\Sigma\subset X$ in
 $M$-theory
on $TN(n) \times X$ where $TN(n)$ is a
(multi-)Taub-NUT
space. Our analysis of section 5 continues
to apply because $TN(n)$ is hyperK\"ahler.
 By holomorphy, only
one orientation of M2 brane can contribute
to $W$. Then, in order to have two
fermion zero-modes, the  Taub-NUT space
must be oriented so that its covariantly constant
spinor is of chirality $\epsilon_-^{ \alpha}$.
Naively the path integral factorizes as the
product of D2 and D6 instanton effects
described below, but we expect D2 D6 interactions
to spoil this.

The superpotential now takes
the form:
\eqn\superpot{
W = \sum_{\Sigma\subset X}
\tau(\Sigma)
e^{i \int_\Sigma (C + i \Phi)}
+ \sum_{TN(n)} Z_n(X)  +{\rm D2D6 \  instantons}
}
where in the first sum $\tau(\Sigma)$ is the path
integral from RW and McLean multiplets,
as described above.

\newsec{A mathematical application: Counting
supersymmetric 3-cycles}

Let us now combine the above results with
mirror symmetry. The family of barely $G_2$
manifolds is preserved
by the mirror map at large complex and
K\"ahler structure.  That is, if $Z$ is a Calabi-Yau
with a real structure
$\sigma_Z$, then for zero axion (i.e.
pure imaginary complexified K\"ahler form)
the mirror $\tilde Z$ has the same properties.
For toric Calabi-Yau
manifolds this
can be proven rather directly by using
using the explicit formulae for periods and
special coordinates given in
\hkty. It also follows from
the gauged linear sigma model approach to
mirror symmetry described in \morrpless.
Physically the assertion is not at all
surprising since the existence of a
real  structure on $Z$
implies  CP invariance of heterotic string
compactified on $Z$ \stromwitten.  We
expect that  if
$\sigma_Z$ has no fixed points then neither does
$\tilde\sigma_{\tilde Z}$ for a reason given below.

Let us now consider two mirror Calabi-Yau
3-folds $Z$ and $\tilde Z$.  A choice of
$Z$ is determined by a choice of complexified
K\"ahler class $B+ i J$ and complex structure, the
latter
encoded in the holomorphic $(3,0)$ form
$\Omega$. If $\tilde \sigma_{\tilde Z}$
acts freely then by
  strong mirror symmetry we have:
\eqn\smsii{
IIA[ { Z(B+i J , \Omega) \times S^1_{R}
\over  (\sigma_Z, -1)} ;g_s ; \ell_s] =
IIA[{ \tilde Z(\tilde B+i \tilde J , \tilde\Omega)\times
S^1_{\tilde R} \over  (\tilde \sigma_{\tilde Z},-1)}
;\tilde g_s ; \ell_s]
}
where the  freely acting anti-holomorphic
involutions $\sigma_Z, \tilde \sigma_{\tilde Z}$
are mapped into each other by mirror symmetry.
Note that if $\sigma_Z$ acts freely then
so should $\tilde \sigma_{\tilde Z}$. For, if $\tilde \sigma_{\tilde Z}$
did not act freely, then the fixed point locus could be
resolved by blowing up $A_1$ singularities as in
\joycei. In that case, there would be an enhanced
nonabelian $SU(2)$ gauge symmetry in the mirror theory.
(We cannot use mirror symmetry to prove
that $\tilde \sigma_{\tilde Z}$ acts freely because
the reasoning would be circular.)

The superpotentials
in the three-dimensional  effective supergravities based on
$Z, \tilde Z$ must be equal. Since we can take
the volumes $V, \tilde V$ to be both large
(by going to both large K\"ahler and large
complex structure
moduli of $Z$) we must have equality for the
sum over 3-cycles. Since K\"ahler and complex structure
moduli are exchanged by mirror symmetry we conclude
that
\eqn\threecyc{
W_{\rm lag}(Z) = W_{\rm hol}(\tilde Z).
}
It is widely expected that  under mirror
symmetry there should be a connection
between holomorphic curves and Lagrangian
3-cycles. This has been discussed
in \kontsevich\ and is closely connected to
the SYZ construction \syz. For further
recent discussion see
\refs{\polishchuk, \thomas, \tyurin}.
As far as we know, the ``explicit'' counting
formula \threecyc\  is new.
(Similar proposals were of course made in
\bbs. These concern the curvature of
hypermultiplet moduli spaces.  In order to
make the proposal of \bbs\ more
explicit we would need to
investigate the one-loop measures
for several different kinds of brane-instantons.)

\newsec{Heterotic Duals}

It is interesting to consider the interpretation
of membrane-induced superpotentials in
the context of dual heterotic models.
Some conditions for the existence of such
dual pairs were investigated in
\refs{\paptown,\hls,\acharyaii,\shatii,\shatiii}. Here we
focus on barely $G_2$ manifolds
$X =(Z \times S^1)/\IZ_2$. If $Z$ admits both
a K3 fibration $p: Z \rightarrow \IP^1$ and
an elliptic fibration with section then
\eqn\duality{
M[Z \times S^1] = IIA[Z] = HET[S_H \times T^2, V]
}
where $S_H$ is a K3 surface and $V $ is
an $E_8 \times E_8$ gauge bundle.
(See \aspinwall\ for a review.)
Taking a further quotient by $\IZ_2$
we may expect $M[X_7] = HET[Z_H, V_H] $
for a Calabi-Yau $Z_H$ and gauge bundle
$V_H$. We can give a heterotic string interpretation
of the membrane instantons $[\Sigma^{\rm hol}]$
and $[\Sigma^{\rm lag}]$ of section seven
by examining the coupling
constant dependence of the instanton action,
as in sec. 4 of \wfv.

Let us consider first $[\Sigma^{\rm hol}]$.
{}From \anothnorm\ we find
\eqn\actone{
\int_{\Sigma^{\rm hol}}  \Phi  = \half (R/\ell)
\biggl( \int_{\Sigma^-_2} J \biggr)
}
where $R$ is the radius of the $M$-theory circle,
and $J$ is the K\"ahler class of $Z$. Making the
standard Weyl rescaling to the IIA string metric
we get the action $\half (\int_{\Sigma^-_2} J )/\ell_{IIA}^2$
where the K\"ahler class is now with
respect to the IIA
string metric and $\ell_{IIA}$ is the string scale.
Now we choose a basis of K\"ahler forms for
$Z$ so that the complexified K\"ahler moduli
$\tau_H , y^i$ correspond under \duality\ to the
heterotic axiodil and Wilson lines.
Under string duality, $\tau_H$ is the
K\"ahler class of a section $\sigma$
of the fibration $\sigma: \IP^1 \rightarrow Z$
\aspinwall.
Let $[\Sigma^i]$ be dual to $y^i$. If
$[\Sigma^-_2] = n_0 [\sigma(\IP^1)] +
n_i [\Sigma^i]$ then the instanton
contributes an effect $\Delta W$ which
depends on vectormultiplet moduli
of $IIA[Z]$ according to:
\eqn\actwo{
\exp\biggl[i  ( n_0   \tau_H + n_i   y^i) \biggr] .
}
If $n_0 =0, n_i \not=0$ we may interpret
these effects in the heterotic theory
as worldsheet instantons localized
in $S_H$. If $n_0 \not=0, n_i = 0 $ we may
interpret these effects as spacetime instantons,
corresponding to small heterotic 5-branes.
If both $n_0, n_i \not=0$ we have mixed
instantons.

Let us now turn to Lagrangian-type cycles
$[\Sigma^{\rm lag}]$. The action only
depends on hypermultiplet
moduli of $IIA[Z]$ and in
IIA units at large volume is given by \bbs:
\eqn\actthree{
\exp\Biggl[ - {1 \over  2 g_s} {\vert \int_\Sigma \Omega \vert \over  \sqrt{i
\int_Z \Omega \wedge \bar \Omega } } + i\int C^{(3)}
\Biggr]
}
Here $g_s$ is the 4-dimensional string coupling
for $IIA[Z]$,
$\Omega$ is a holomorphic $(3,0)$ form on
$Z$ and $C^{(3)}$ is now a RR potential.
Further orbifolding by $\IZ_2$ should not
introduce dependence on the heterotic
axiodil (part of a vectormultiplet) and hence
we conclude that these instanton effects
correspond to heterotic worldsheet
instantons.

This discussion raises several questions.
First, it would be nice to understand how
the expression \actthree\ becomes a holomorphic
function of the complex structure and bundle
moduli of  $[Z_H,V_H]$.
(This is partially answered in \bbii.)
Second, it is interesting
to note that in the gauged linear sigma model approach
to $(0,2)$ models \ewglsm\distlerkachru\ the
complexified K\"ahler moduli have a very
different origin (FI terms) from the complex structure
and bundle moduli (superpotential terms). It would
be interesting to see if this distinction is
related to the above distinction of types of
membrane instantons. Finally, we note that,
as in \wfv,
 it follows that world-sheet instantons
can indeed destabilize $(0,2)$ theories.
It should be interesting to reconcile \superp\
with the phenomena discussed in \silverwitten.

\newsec{Open membrane instantons}

Another interesting
application of our results is to open
membrane  instanton effects.
These effects are not well-understood,
but should be important in learning about
mirror symmetry through the SYZ construction \syz,   the
physics of the D1D5 system
(via the papers \ghm\swrec), and  in obtaining a
deeper     understanding of $(0,4)$ models
of supersymmetric black holes \msw\mmt. They should
also play an important  role in the models
of low energy physics discussed in
\refs{\horavawitten, \wittsc,\bd, \horava,\ovrutii,\ovrut}.
There are many nontrivial issues one
must discuss in order to do detailed computations.
We hope to return to these elsewhere and
limit ourselves here to some brief preliminary
discussion.

We consider
 the strongly coupled $E_8 \times E_8$
heterotic string, realized as $M$-theory
on $\IR^4 \times S^1/\IZ_2 \times Z$. There can
also be 5-branes wrapping $\IR^4 \times \{x\} \times S$,
where $S\subset Z$ is a holomorphic curve.
There are now
several kinds of membrane instantons.
In addition to the closed 3-manifolds
discussed thus far in this paper there are
membranes ending on  5-branes and/or
  9-branes.  The general
form of the superpotential will be
\eqn\genform{
W = W_c + W_{55} + W_{59} + W_{99} + W_{gc}
}
Here $W_c$ is the superpotential
arising from closed membrane instantons.
  $W_{gc}$ is the superpotential generated
by strong infrared dynamics of
   the boundary gauge
theories  ( discussed in \bd\horava\ovrutii).
$W_{99}$ correspond to the worldsheet
instantons \horavawitten.

Let us now consider briefly open membranes
ending on 5branes.
\foot{ References to the substantial literature
on this subject will be given in the hypothetical
future paper mentioned above. }
The physical ``gerbe connection'' on the 5brane
is $C- d\beta$ where $\beta$ is the chiral 2-form on the
5-brane worldvolume. It follows that the instanton
action for a membrane ending on a 5brane is given
by
\eqn\membweight{
q_{\rm membrane} :=
\exp\biggl[ i\bigl( \int_{\p \Sigma} \beta + i \int_\Sigma \ell^{-3} \vol
\bigr)\biggr]
}
The parameter \membweight\
 will play the role for membranes analogous
to the complexified K\"ahler expansion parameter
for open worldsheet  instantons. In particular,
consider
the term $W_{55}$ corresponding  to membranes
stretching  between different 5-branes.
Reduction of the chiral 2-form  on each 5brane
results in a complex scalar  $a_i + i x_i$
in a chiral multiplet, where
  $x_i$ is the position of the
5-branes on the interval $S^1/\IZ_2$
The superpotential $W_{55}$ will come from
membrane instantons stretching between
holomorphic curves in the 5brane worldvolumes.
These will have  the topology
$\Sigma= S \times [0,1]$ and therefore contribute
\eqn\superff{
W_{55}  =
\sum_{x_i > x_j }
e^{ [a_i - a_j  + i (x_i - x_j)] (z \cdot S) }
t_{55}^{(ij)}
}
Here $z \cdot S $ is the chiral modulus
associated with the size of the holomorphic
cycle $S\subset Z $ on which the
membrane ends,
$ z \cdot S:=   \int_{S} (B+ i J)$,
where $B  = \iota({\p \over  \p x} ) C $ is
the superpartner of $J$. We also
expect that the coefficients
$t_{55}^{(ij)} $ will be essentially
given by transition amplitudes in the RW and
McLean topological field theories
discussed above.

The   expression \superff\
only holds for $x_i - x_j \gg \ell$.
This raises the interesting question of
holomorphy since for the other order
the instanton action
must be suppressed by
$\sim \exp[- \vert x_i - x_j \vert] $. Thus there
   appears
to be nonanalytic behavior on a real codimension one
locus. The resolution of this problem is probably
that there is a nontrivial multiple cover formula
for the 55 instantons. Thus, we expect a
series in $q$ defined by \membweight\
and different power series expansions
dominate in the regions
$x_i - x_j \gg \ell$ and $x_j - x_i \gg \ell$.
Note that this implies there should be a pole in
$W_{55}$ so it certainly cannot always be
neglected relative to other terms (like $W_{gc}$).

Similar considerations
hold for $W_{59}$. Toy models
(i.e., choosing a simple ansatz for the
K\"ahler potential)  indicate
that one can generate interesting potentials
for the 5-brane position moduli.

\newsec{Conclusions}

\subsec{Applications}

One immediate application of the above
results is that certain $G_2$ compactifications
of $M$-theory are unstable quantum mechanically in the regime
where our calculation is valid, that is for  3-cycles
whose volume is large compared to the
Planck scale. This class includes smooth $G_2$-manifolds
which admit susy 3-cycles which are rational
homology spheres.
Note that we do not expect any further corrections
to the superpotential from non-trivial low-energy dynamics
since the low-energy gauge theory at generic points in the
moduli space consists of $U(1)$ gauge theory with non-chiral
matter.
Thus, (as is hardly surprising),
there is an M-theoretic Dine-Seiberg
problem \dineseib.

\subsec{Potential applications}

We   think the mathematical applications
for enumerating supersymmetric 3-cycles
in Calabi-Yau 3-folds are
very promising. We anticipate some
fairly amazing identities for sums over supersymmetric
3-cycles weighted by topological invariants
such as Ray-Singer torsion and the Casson invariant.

We hope that some of the technical results
found above will be useful in further investigations
of brane-induced instanton effects in
other compactifications.
As one example, it should be possible to be more
explicit about the nonperturbative corrections to
hypermultiplet geometry discussed in  \bbs.
It follows, for example, that for a rigid
special Lagrangian 3-cycle $\Sigma$ the factor
$N$ in \bbs\  eq. 2.49 is just the order of the
finite group:
$\vert H_1(\Sigma;\IZ)\vert $.

Our methods suggest some interesting
formulae for instanton/anti-instanton
effects such as $D2 \overline{D6}$ effects.
Indeed, it follows from \rw\ that such  effects
will involve the Casson invariant of $\Sigma$.
These effects contribute to the induced
  {\it potential}. On the other hand, we know
from supergravity that
\eqn\potential{
V = e^K \biggl( \vert\CD W\vert^2 - 3 \vert W \vert^2
\biggr)
}
 It is possible that by comparing expressions
we will learn something about quantum corrections
to the {\it K\"ahler potential} in $d=4, \CN=1$
$M$-theory compactifications.
Of course, this requires first an understanding
of perturbative contributions to $K$.
In any case, an
understanding of the K\"ahler potential
is essential to addressing questions of
supersymmetry breaking, and is sadly lacking.

\subsec{Some Unfinished Business}

The present paper has, regrettably,
a somewhat programmatic flavor.
Several points related to the above
discussion should be looked into much more
thoroughly than we have done here.
First, and foremost,
the rules for systematic computation
of membrane instanton effects should be
clarified.

Second,
we determined $W$ through the chiral multiplet
mass matrix. When one considers the
gravitino and gaugino mass matrices one
encounters important contact interactions.
These need to be sorted out. It is quite likely
that doing so will shed light on nonperturbative
corrections to the gauge kinetic function
$\tau_{IJ}(z)$.

Third, we have only stated a complete claim for
rigid membrane instantons. In general
there might be contributions from \
 nonisolated instantons. The
case of worldsheet instantons can give us
some guidance, but the full story remains to
be understood. Similarly, the effects of
open membrane instantons are potentially
very important, but much work remains to
be done here.

\appendix{A}{Conventions and notations}

\subsec{Index conventions}

We use AMS fonts for superspace indices and
for superfields (e.g.$ \bbz^{\bbm}$).
 $M=0,\dots, 10$ is a bosonic world index
and $\mu=1,\dots 32$ is a fermionic world index.
(We also use $\mu=0,1,2,3$ for bosonic
world indices in
the noncompact spacetime $M_4$, but the
distinction is clear from context.)
$A, \alpha$ are the tangent frame counterparts
of $M,\mu$.
$e_M^{~~A}$ is a frame for the metric at $\Theta=0$.
$\Gamma_{MN\dots}$ are antisymmetrized gamma
matrices of weight one. Worldvolume indices on
$\Sigma$ are denoted by $1\leq i,j,k \leq 3$,
and a generic worldvolume coordinate system
is denoted $s^i$.
Our Lorentzian signature is mostly $+$.

\subsec{Differential geometry}

Our conventions for
differential forms are
$C  = {1 \over  3!} C_{MNP} dx^M dx^N dx^P$.
Our normalization of the second fundamental
form is:
\eqn\secff{
\Gamma^{m''}_{k' l' } = - \half h^{m'' n''} Q_{k' l' n''}
}
where $\Gamma^{m''}_{k' l' } $ is the Christoffel
connection in the ambient space, restricted to
$\Sigma$. Real harmonic $p$-forms on a manifold $\CM$
are denoted $\CH^p(\CM;\IR)$. The forms corresponding
to integral classes under the Hodge-DeRham
isomorphism are $\CH^p(\CM;\IZ)$.

If $V$ is a
vector bundle with metric we let $\CS(V)$
(or $\CS^\pm(V)$) denote the minimal dimension associated
spin bundle (there is an implicit choice of
spin structure).

We   often abbreviate
Calabi-Yau to CY.

\subsec{Some spinor conventions  }

In section four we use the following Clifford
algebra conventions. The three-dimensional Clifford algebra on the
M2 world-volume is
\eqn\threedmink{
\eqalign{
\tau^0 = i \sigma^2 \qquad \tau^1 = \sigma^1 \qquad
& \qquad \tau^2 = \sigma^3 \cr}
}
The orientation is
$\epsilon_{012}   = - \epsilon^{012}  = +1 $.
The Euclidean continuation is $\tau^3 = \sigma^2$.

The 7-dimensional
Euclidean Clifford algebra $C\ell_7$ is nicely
realized using the octonions $\bbo$.
In the Cayley-Dickson description
the octonions can be thought of as
pairs of quaternions  with  multiplication
\eqn\octmult{
(a,b) \cdot (c,d) \equiv (ac - \bar d b, d a + b \bar c).
}
Choosing an isomorphism of $\im \bbo\cong \IR^7$
we define Clifford multiplication by an
orthonormal  basis
to be octonionic multiplication by the imaginary
units. This gives a representation of $\gamma_i$ by
$8\times 8 $ real antisymmetric matrices.

The Clifford algebra $C\ell_8$, used for the
normal directions to the membrane, is
represented by
\eqn\eightdeuc{
\eqalign{
\Sigma^{1,\dots, 8} & = \pmatrix{ 0 & \gamma^{1,\dots, 8}\cr
\tilde \gamma^{1,\dots, 8} & 0 \cr} \cr
\tilde \gamma^{1,\dots, 7 } & = - \gamma^{1,\dots, 7 }\cr
\tilde \gamma^{8 } & = \gamma^{8 }\cr}
}
where $\gamma^i$ is the representation of
$C\ell_7$ described above, and
$\tilde \gamma^{1,\dots, 7}$ is  the other inequivalent
representation of $C\ell_7$.  We take
 $\gamma^8_{a \dot a} = \delta_{a \dot a}$. The chirality operator is
\eqn\chiral{
\bar \Sigma   = \pmatrix{ -1_8 & 0 \cr 0 & 1_8\cr}
}
Note that $(\Sigma^i)^{T} = \Sigma^i$, and
$ (\Sigma^i)^{*} = \Sigma^i$, $i=1,\dots, 8$.

We take the  representation of the Clifford
algebra $C\ell(1,10)$ to be
\eqn\eightdeuc{
\eqalign{
\Gamma^{m'} & = \tau^{m'} \otimes \bar \Sigma \cr
\Gamma^{m''} & = 1_2 \otimes \Sigma^{m''} \cr}
}
It is often convenient to introduce
spinor index notation: $A=1,2$, $a,\dot a=1,\dots, 8$.
The above matrices are written
\eqn\spinindx{
\eqalign{
(\tau^{m'})^A_{~~B} \qquad & \qquad \gamma^{m''}_{a \dot a} ,
\tilde \gamma^{m''}_{\dot a , a} \cr}
}
Note that
$\tau^{m'}_{AB} := (\tau^0 \tau^{m'} )^A_{~~B}
 = \tau^{m'}_{BA}$
is symmetric. We raise/lower spinor
indices with $\varepsilon^{AB} = (i \sigma^2)^{AB}$.
The Euclidean continuation is obtained by $\tau^3 = \sigma^2$.
Then the charge conjugation matrix
$(\Gamma^M)^T = - C \Gamma^M C^{-1}$ is
given by $C = i \sigma^2 \otimes \bar \Sigma$.

\appendix{B}{Scales and dimensions}

Our conventions on units and scales are the following:
$\ell$ is the 11-dimensional Planck length.
$g_{MN}= \eta_{MN} + h_{MN}$ is dimensionless.
Local coordinate
differentials $dx^M$ have dimensions of length.
Thus $[\vol(g)]   = L^{d}$ on a $d$-dimensional
manifold.
A field (like $C$ ) which is a differential form  is
dimensionless.  Thus
\eqn\cee{
C = {1 \over  3!} C_{MNP} dx^M dx^N dx^P
}
means $C_{MNP}$ has  dimensions $L^{-3}$.
The gravitino $\Psi_M$ has dimension
$[\Psi_M] = L^{-1/2}$.
  With these conventions, the Hodge star $*$
acting: $\Omega^k \rightarrow \Omega^{n-k}$
has dimensions of $L^{n-2k}$.
Since the components of    $\Phi_{mnp}$ are,
in an orthonormal  frame,   the  dimensionless
structure constants of the octonions we have
$[\Phi]   = L^3$.

Our normalization of the 11D supergravity
action is nonstandard. To translate to the
normalizations used, for example, in
\horavawitten\  one sets:
\eqn\maphw{
\eqalign{
C^{\rm here}_{MNP} & = {3 \sqrt{2} \over  \ell^3} C^{HW}_{MNP}\cr
(4 \pi)^2 \ell^9 & = \kappa^2\cr}
}
Also, note that we use standard normalizations for
differential forms so $(d C)_{MNPQ} = \p_M C_{NPQ} + 3 terms$. Finally we made
a constant rescaling
by $g \rightarrow 2^{2/3}  g$.

\appendix{C}{Some explicit susy 3-cycles}

One well-known way to find supersymmetric
3-cycles is as the fixed point set of a real
structure in a Calabi-Yau manifold $Z$.
At first sight it appears that one cannot
use this idea  to construct susy
3-cycles in   barely $G_2$ manifolds, because
$X_7$ is only smooth if $\sigma_Z$
acts without fixed points - but then we
lose the susy cycle!  Of course,
a CY can have
several different real structures.
Any two differ by a   holomorphic isometry.
Therefore, using CY's with both real structures
and symmetries we can use one real
structure $\sigma_Z$ (fixed point free)
to make a barely $G_2$ manifold and study
the fixed points of another real structure
$\tilde \sigma_Z$ to find susy 3-cycles.

We now give an explicit example of this technique.
We consider the family in $\IP^5[2,4]$:
\eqn\truexpl{
\eqalign{
\sum_{i=1}^6 X_i^2 & = 0 \cr
\sum_{i=1}^6 a_i X_i^4 & = 0 \cr}
}
We assume $a_i\not=0$.
The discriminant locus is then defined by the
equations:
\eqn\disclc{
\sum_{i\in I_\alpha} {1 \over  a_i} = 0
}
where $I_\alpha$ are subsets
of $\{1,2, \dots, 6\}$ with 2 or more elements.
We will take $a_i$ so that the CY is smooth.
without loss of generality
 we may take $a_2 + a_3 =1$.

If $a_i$ are real we have the
  standard real structure   $\sigma_Z: X_i \rightarrow
\bar X_i$. The fixed points are all real, and there
are clearly no real solutions to \truexpl.  Thus we
can build a smooth barely $G_2$ manifold.
We will study the alternative real
structure with $\tilde \sigma_Z(X_i) = \bar X_i$, $i=1,2$
and
$\tilde \sigma_Z(X_i) =- \bar X_i$, $i=3,\dots, 6.$
 The fixed point locus is:
\eqn\altstrct{
\eqalign{
X_{1,2} & = Y_{1,2} \cr
X_{3,4,5,6} & = i Y_{3,4,5,6} \cr}
}
where the coordinates  $Y_i$ are   real.
Thus, the supersymmetric cycle is the intersection
of the fixed point locus \altstrct\ and \truexpl\
and is given by the solutions with $Y_i $ real to
\eqn\realcycl{
\eqalign{
Y_1^2 + Y_2^2 & = Y_3^2 + r^2\cr
a_1 Y_1^4 + a_2 Y_2^4 + a_3 Y_3^4 + f_4 & = 0 \cr}
}
Here
 $r^2 = Y_4^2 + Y_5^2 +Y_6^2$
and $f_4$ is    the  quartic defined above.
It is   useful to define: $f_4 := r^4 \bar f_4$, where
$\bar f_4$ is just a function of the polar angles.
We assume $a_2 + \bar f_4 >0$ for all angles.

Now, it is not difficult to see that
the region:
\eqn\solveii{
(a_1 + a_2(r^2-1)^2 + f_4)< 0
}
is a connected bounded region $B_*$
near the origin of $\IR^3$ and diffeomorphic
to $B^3$, the 3-ball.
The boundary of this region
is  the surface $\Sigma_*$ diffeomorphic
to $S^2$.
In the region $B_*$ there are only two real roots for
$Y_3$. These two real roots
collide and vanish along the surface $\Sigma_*$.
Moreover, if $2 a_2>1$
then the two roots for $Y_2$ are both real and
never vanish in the region $B_*$.

We are now in a position to describe our real solution
set. There are 4 copies of the ball $B_*$ labelled by
  the 4 real roots $(\pm Y_2, \pm Y_3)$. Along the
surface $\Sigma_*$ the two roots $Y_3$ vanish,
but the roots $\pm Y_2$ are bounded away from
zero. Therefore, for each root of $Y_2$, we
glue two copies of $B_*$ together along $\Sigma_*$.
We get, topologically, two disjoint copies of
$S^3$.
Now, when we take the quotient by $\sigma_Z$
this acts by taking $(Y_1, Y_2) \rightarrow (Y_1, Y_2)$,
but $(Y_3, \dots, Y_6 )
\rightarrow -(Y_3, \dots, Y_6 )$. This is clearly
homotopic to the antipodal map on $S^3$ acting
separately for each disjoint copy.
We thus get two disjoint copies of
$\IR P^3$ in the barely $G_2$ manifold.
These are rigid   Lagrangian 3-cycles.

\bigskip
\centerline{\bf Acknowledgements}\nobreak
\bigskip

Some of these results were presented in the
 Penn Math-Physics seminar Nov. 4, 1998 and again
at the conference ``New ideas in particle physics
and cosmology,'' at the  University of Pennsylvania,
May 21, 1999. GM would like to thank the
organizing comittee for the invitation to speak.
We would  like to
thank  M. Mari\~no, R. Minasian, D. Morrison,
T. Pantev, N. Seiberg,
E. Silverstein, I. Singer, A. Strominger,
and  E. Witten for discussions. We are
particularly grateful to E. Witten for important
remarks on and corrections to a preliminary version
of this paper. GM and JH
would like to acknowledge the hospitality of
the Aspen Center for Physics
and GM would like to thank the Institute
for Advanced Study for hospitality and
the Monell foundation for support.
The  work  of JH is supported by
NSF Grant No.~PHY 9600697,   GM is
supported by
DOE grant DE-FG02-92ER40704.

\listrefs

\bye